%% file: afb_preprint_v6.1.tex
\RequirePackage{lineno}
\documentclass[aps,prd,twocolumn,showpacs,superscriptaddress,floatfix,nofootinbib]{revtex4} 

\usepackage{amsmath}
\usepackage{amssymb}   
\usepackage{graphicx}
\usepackage{bm}
\usepackage{epsfig}
\usepackage{multirow}
\usepackage{url}
\usepackage{array} 
\usepackage{bold-extra}
\usepackage{scalefnt}
\usepackage{hyperref}
\usepackage[shortcuts]{extdash}            
\usepackage[htt]{hyphenat}
\usepackage[caption=false,font=small]{subfig} 

\newcommand*\patchAmsMathEnvironmentForLineno[1]{%
  \expandafter\let\csname old#1\expandafter\endcsname\csname #1\endcsname
  \expandafter\let\csname oldend#1\expandafter\endcsname\csname end#1\endcsname
  \renewenvironment{#1}%
                   {\linenomath\csname old#1\endcsname}%
                   {\csname oldend#1\endcsname\endlinenomath}}%
\newcommand*\patchBothAmsMathEnvironmentsForLineno[1]{%
  \patchAmsMathEnvironmentForLineno{#1}%
  \patchAmsMathEnvironmentForLineno{#1*}}%
\AtBeginDocument{%
  \patchBothAmsMathEnvironmentsForLineno{equation}%
  \patchBothAmsMathEnvironmentsForLineno{align}%
  \patchBothAmsMathEnvironmentsForLineno{flalign}%
  \patchBothAmsMathEnvironmentsForLineno{alignat}%
  \patchBothAmsMathEnvironmentsForLineno{gather}%
  \patchBothAmsMathEnvironmentsForLineno{multline}%
}

\newcommand \smaller[2][0.72]{{\scalefont{#1}#2}} 

\newcommand {\this}	{paper}

\newcommand {\plainafb}	{\ensuremath{A_{\mathrm{FB}}}}
\newcommand {\afb}	{{\plainafb}}

\newcommand {\pbar}	{\ensuremath{{\bar p}}}
\newcommand {\ppbar}	{\ensuremath{{p\pbar}}}
\newcommand {\qbar}     {\ensuremath{{\bar q}}}
\newcommand {\bbar}     {\ensuremath{{\bar b}}}
\newcommand {\tbar}     {\ensuremath{{\bar t}}}

\newcommand {\ttbar}    {\ensuremath{{t\tbar}}}
\newcommand {\pjets}    {{\textrm{+jets}}}
\newcommand {\wpj}      {{\ensuremath{W\!+}jets}}
\newcommand {\lpj}      {{\ensuremath{l\pjets}}}

\newcommand {\lptj}      {{\ensuremath{l\textrm{+}3\textrm{\,jet}}}}
\newcommand {\lpgefj}      {{\ensuremath{l}+\ensuremath{\geq}4\,jet}}
\newcommand {\lpgetj}      {{\ensuremath{l}+\ensuremath{\geq}3\,jet}}
\newcommand {\threej}      {{3 jets}}
\newcommand {\gefj}      {{\ensuremath{\geq}4 jets}}
\newcommand {\getj}      {{\ensuremath{\geq}3 jets}}
\newcommand {\geob}      {{$\geq$1 $b$ tags}}
\newcommand {\getb}      {{$\geq$2 $b$ tags}}
\newcommand {\oneb}      {{$1$ $b$ tag}}

\renewcommand {\mtt}   {\ensuremath{{m_\ttbar}}}

\newcommand {\GeV}        {\ensuremath{\,\textrm{GeV}}}
\newcommand {\TeV}        {\ensuremath{\,\textrm{TeV}}}

\newcommand {\ifb}        {\ensuremath{\,\textrm{fb}^{-1}}}

\newcommand {\pt}         {{\ensuremath{p_T}}}
\newcommand {\dy}         {{\ensuremath{\Delta y}}} 
\newcommand {\absdy}      {{\ensuremath{\left|\dy\right|}}}

\newcommand {\djom}  {{\ensuremath{\Delta \phi (\mathrm{jet}_1,\met)}}}

\newcommand{\emiss}{/\!\!\!\!E}
\newcommand{\met}{\ensuremath{{\emiss_\textrm{T}}}}
\newcommand{\lbpt}	{\ensuremath{{p^{\textrm{LB}}_{T}}}}
\newcommand{\ktmin}	{\ensuremath{{k^{\textrm{min}}_T}}}
\newcommand{\mjj}	{\ensuremath{{M_{jj}}}}
\newcommand{\chisq}  {{\ensuremath{\chi^2}}}
\newcommand{\ptthree}   {{\ensuremath{p_T^\mathrm{3rd}}}}
\newcommand{\mjjmin}   {{\ensuremath{M_{jj}^\mathrm{min}}}}
\newcommand{\mvab} {{\ensuremath{\textrm{MVA}_b}}}
\newcommand{\pcor} {{\ensuremath{P_c}}}

\newcommand {\migaccmat}  {\ensuremath{\boldsymbol{T}}}
\newcommand {\migmat}  {\ensuremath{\boldsymbol{M}}}
\newcommand {\accmat}  {\ensuremath{\boldsymbol{A}}}
\newcommand {\regmat}  {\ensuremath{\boldsymbol{R}}}
\newcommand {\covmat}  {\ensuremath{\boldsymbol{\Sigma}}}

\newcommand {\mttbar} {{\ensuremath{m_\ttbar}}}

\newcommand {\Nf}         {\ensuremath{{N_f}}}
\newcommand {\Nb}         {\ensuremath{{N_b}}}
\newcommand {\Ntag}         {\ensuremath{{N_\textrm{tag}}}}

\newcommand {\Vrr}  {{\ensuremath{\boldsymbol{V}}}}

\newcommand {\aady} {{\ensuremath{\afb\left(\absdy\right)}}}

\newcommand {\DZ}         {{D0}} 

\newcommand {\noone} {\hphantom{\ensuremath{1}}}

 \newcommand {\mcatnlo}  {\smaller{MC@NLO}}
 \newcommand {\alpgen}   {\smaller{ALPGEN}}
 \newcommand {\herwig}   {\smaller{HERWIG}}
 \newcommand {\madgraph}   {\smaller{MADGRAPH}}
 \newcommand {\pythia}   {\smaller{PYTHIA}}
 \newcommand {\comphep}   {\smaller{COMPHEP}}
 \newcommand {\cteq}   {\smaller{CTEQ6.1}}
 \newcommand {\mrst}   {\smaller{MRST2003}}
 \newcommand {\geant}    {\smaller{GEANT}}
 \newcommand {\tunfold}   {{TU\smaller{NFOLD}}}
 \newcommand {\prodhlvl} {production\-/level}
 \newcommand {\Prodhlvl} {Production\-/level}
 \newcommand {\recohlvl} {reconstruction\-/level}
 \newcommand {\Recohlvl} {Reconstruction\-/level}

 \newcommand {\eg}       {{\textrm{e.g.}}}
 \newcommand {\etal}     {{\textit{et al.}}}

\newcommand {\bsd} {background\hyp{}subtracted data}

\newcommand {\tstrut} {\rule{0pt}{2.5ex}}
\newcommand{\centercell}[1]{\multicolumn{1}{c}{#1}}
\newcommand{\head}[1]{\centercell{#1}} 
\newcommand{\multihead}[2]{\multicolumn{#1}{c}{#2}}

\newcommand {\dvmLegendBoth}{The left column shows \lptj\ events, and the right column  shows \lpgefj\ events. 
Rows from top to bottom display events with 0, 1, and $\ge$2 $b$ tags.}
\newcommand {\dvmLowerPanel}{The ratio between the data counts and the model expectation is shown in the lower panel, with
the hashed area representing the systematic uncertainties.}
\newcommand {\dvmLowerPanels}{The ratio between the data counts and the model expectation is shown in the lower panels, with
the hashed area representing the systematic uncertainties.}

\makeatletter
\DeclareRobustCommand*{\bfseries}{%
  \not@math@alphabet\bfseries\mathbf
  \fontseries\bfdefault\selectfont
  \boldmath
}
\makeatother

\DeclareMathOperator{\erf}{erf}

\begin{document}
\title{Measurement of the forward--backward asymmetry in   \\
top quark--antiquark production in $\ppbar$ collisions using the lepton+jets channel}

\input author_list_revised.tex

\date{April 30, 2014}
           
\begin{abstract}
We present a measurement of the forward--backward asymmetry in top quark--antiquark production
using the full Tevatron Run II dataset collected by the D0 experiment at Fermilab.
The measurement is performed in lepton+jets final states 
using a new kinematic fitting algorithm for events with four or more jets and
a new partial reconstruction algorithm for events with only three jets.
Corrected for detector acceptance and resolution effects, 
the asymmetry is evaluated to be $\afb = \left(10.6 \pm 3.0 \right) \%$. 
Results are consistent with the standard model predictions which range from 5.0\% to 8.8\%. 
We also present the dependence of the asymmetry on the invariant mass of the top quark--antiquark system
and the difference in rapidities of top quark and antiquark. 
\end{abstract}


\maketitle

\section{Introduction}
\subsection{Motivation and definitions}
\label{sec:intro}
Over the last five years both  experiments at the Fermilab Tevatron Collider measured positive forward--backward asymmetries 
in the production of top quark--antiquark pairs in proton--antiproton collisions ($\ppbar \rightarrow \ttbar$)~\cite{bib:p17PRL, bib:CDFPRL, bib:CDFdep, bib:ourPRD, bib:CDF2012}. 
The reported values were consistently above predictions of the standard model of particle physics (SM)~\cite{bib:KnR, bib:other}. 
In particular, the CDF Collaboration observed a strong rise of the asymmetry with the 
invariant mass of the \ttbar\ system, \mttbar~\cite{bib:CDFdep}. 
The dependence of the asymmetry on \mttbar\ in \DZ\ data, as measured  in Ref.~\cite{bib:ourPRD},
was statistically compatible with both the SM predictions and with the CDF result. 
Several beyond-the-SM scenarios were suggested to explain the measured \afb\ values~\cite{bib:BSM},
in particular using the framework of parity-violating strong interactions suggested in
Ref.~\cite{bib:axigluon}. In this \this\ we report new results
from the \DZ\ experiment based on the full dataset collected during Run II of the Fermilab Tevatron Collider, which supersede the result of Ref.~\cite{bib:ourPRD}.

In proton--antiproton collisions, top quark--antiquark pairs are predominantly produced via valence quark--antiquark annihilation. 
Thus, the direction of the proton (antiproton) almost always coincides with the direction of the incoming quark (antiquark). 
We define the difference in 
rapidity\footnote{
The rapidity $y$ is defined as 
$y=\frac{1}{2}\ln\left[\left(E+p_z\right)/
\left(E-p_z\right)\right]$,
where $E$ is the particle's energy and $p_z$ is  its momentum along the $z$-axis, which 
 corresponds to the direction of the incoming proton.}
between the top quark ($y_t$) and antiquark ($y_\tbar$): 
\begin{equation}
\dy = y_t - y_\tbar.
\label{eq:dy}
\end{equation}
We refer to the events that  have $\dy > 0$ as ``forward'', and to those with  $\dy < 0$ as ``backward''.
The forward--backward asymmetry in \ttbar\ production is defined as
\begin{equation}
\afb = \frac {\Nf - \Nb} {\Nf + \Nb},
\label{eq:afb}
\end{equation}
where \Nf\ (\Nb) is the number of  forward (backward) events.
All  \ttbar\ asymmetries reported in this \this\ are given after subtracting the contributions from background processes.

The rapidities of the $t$ and \tbar\ quarks and the corresponding asymmetries can be defined at the production 
level (sometimes denoted as generator level or parton level), 
when the kinematic parameters of the generated top quarks are used. 
Unless stated otherwise the production-level asymmetries are defined for all signal events without 
imposing the selection criteria of this analysis.
The rapidities and asymmetry can also be defined at the reconstruction level, using the reconstructed kinematics of the selected events.
Similarly, the invariant mass of the \ttbar\ system  can be defined at the production and
reconstruction levels.

%
\subsection{ Strategy}
\label{sec:Stra}
In the SM a top quark almost always decays to a $b$ quark and a $W$ boson, which decays either leptonically or hadronically. 
In this \this\ we identify \ttbar\ events using the $\ttbar\to W^+b W^-\bbar$; $W^+ \to l^+ \nu_l$;  $W^- \to q\qbar'$ 
(and charge conjugates) decay chain. 
This channel is commonly referred to as the ``lepton+jets'' (\lpj) channel.  
We select events that contain one isolated lepton (electron or muon) of high transverse momentum (\pt)
and at least three jets. 
The electric charge of the lepton identifies the electric charge of the leptonically decaying $W$ boson and its parent top quark. 
The other top quark is assumed to have the opposite charge. 
The event selection, sample composition determination, and modeling of the signal and background processes are identical
to those used in the measurement of the leptonic asymmetry in \ttbar\ production in the \lpj\ channel~\cite{bib:our_afbl}.
The four-vectors of the top quarks and antiquarks in the events containing at least four jets are reconstructed with a 
kinematic fitting algorithm, 
while  for the events that contain only three jets a partial reconstruction algorithm is used.
If a jet exhibits properties consistent with a jet originating from a $b$ quark,
such as the presence of a reconstructed secondary vertex, we identify it as a $b$-tagged jet~\cite{bib:btagging}. 
The \lpj\ events are separated into
channels defined by jet and $b$-tag multiplicities. 
The amount of signal and the forward--backward asymmetry at the reconstruction level are determined using a 
simultaneous fit to a kinematic discriminant in these channels. 

The measured background\-/subtracted one\-/dimensional (1D) distribution in \dy\ is corrected to the production level (``unfolded'').
From this distribution we calculate the fully\-/corrected \afb\ as well as its dependence on \dy.
To study the dependence of the asymmetry on the invariant mass of the \ttbar\ system, 
 unfolding is done on the background-subtracted data distributions in two dimensions (2D: \dy\ vs \mtt). 
The signal channels are unfolded simultaneously to yield the desired 1D or 2D
\prodhlvl\ distributions, from which the \prodhlvl\
\afb\ values are computed  using Eq.~\ref{eq:afb}. The procedure is
calibrated using simulated samples with varied asymmetries and input distributions in \dy\ and \mtt. 
The statistical and systematic uncertainties of the results
are evaluated using ensembles of simulated pseudo\-/datasets (PDs). 

%
%
\section{D0 detector}
\label{sec:D0}
We use the data collected by the D0 detector during Run II of the 
Tevatron in the years 2001--2011. After imposing event quality requirements ensuring
that all detector systems were fully operational, this dataset corresponds to an integrated luminosity of
$9.7\ifb$. The \DZ\ detector is described in detail elsewhere~\cite{bib:D0det}. 
The central   tracking system, consisting of a silicon microstrip tracker and a scintillating fiber tracker, is  
enclosed within a $1.9\,$T superconducting solenoid magnet. 
Tracks of charged particles are reconstructed within a detector pseudorapidity region\footnote{The detector pseudorapidity 
$\eta$ is defined as $-\ln\left[\tan\left(\frac{\theta}{2}\right)\right]$, 
where $\theta$ is the polar angle measured with respect to the center of the detector. 
The angle $\theta=0$ corresponds to the direction of the incoming proton.} of $|\eta|<2.5$.
Electrons, photons, and jets of hadrons are identified~\cite{bib:ehadID} using a liquid-argon and uranium-plate calorimeter, 
which consists of a central barrel covering up to $|\eta|\approx1.1$, 
and two endcap sections that extend coverage to $|\eta|\approx4.2$~\cite{bib:run1det}. 
Central and forward preshower detectors are positioned in front of the corresponding sections of the calorimeter. 
A muon system consisting of layers of tracking detectors and scintillation counters placed in front of and behind 
 $1.8\,$T iron toroids~\cite{bib:run2muon} identifies muons~\cite{bib:muID} within $|\eta|<2$. 
Luminosity is measured using arrays of plastic scintillators located in front of the endcap calorimeter cryostats.
A three-level trigger system selects interesting events at the rate of $\approx 200\,$Hz for offline analysis~\cite{bib:D0trig}.

%
%
\section{Event  selection and modeling}
\label{sec:selection}
Object reconstruction and identification, as well as event selection, are the same as in Ref.~\cite{bib:our_afbl} and
 are briefly outlined in this section.
We select events with exactly one isolated electron within the detector pseudorapidity range of $|\eta|<1.1$ or one isolated muon within $|\eta|<2.0$, 
and at least three jets within $|\eta|<2.5$.  To limit the possible contribution of poorly modeled background due to multijet production,  leptons of either flavor are required to have  $|y|<1.5$. 
The presence of a neutrino is inferred from a transverse momentum imbalance, which is measured primarily using calorimetry
and is referred to as the ``missing transverse energy'' \met.
All selected objects are required to have transverse momentum $p_T>20\,$GeV, and the jet with the largest $p_T$ (the leading jet) is also required to have $p_T>40\,$GeV.

To identify jets that are likely to be associated with $b$ quarks, we perform a multivariate analysis (MVA) that combines variables characterizing 
the properties of secondary vertices and of tracks with large impact parameters
relative to the primary \ppbar\ interaction vertex (PV)~\cite{bib:btagging}.  The output of the MVA is a continuous variable, \mvab. 
The requirement on \mvab\ ($b$ tagging) used in this analysis has an efficiency of about 64\%
for identifying $b$ jets originating from top quark decay,
and a misidentification probability of about 7\% for jets that do not contain heavy flavor quarks and are produced in  association 
with leptonically decaying $W$ bosons.

We simulate \ttbar\ production using  \mcatnlo\ program (version 3.4)~\cite{bib:mcatnlo}
 with the parton showering performed by \herwig~\cite{bib:herwig}. 
This simulation is fully integrated with the D0 software, allowing for detailed studies of the kinematic 
dependences of \afb\ and their interplay with selection and reconstruction effects.
The main source of background to the \ttbar\ signal is the production of a leptonically decaying $W$ boson in association with jets (\wpj).
The kinematic properties of this process are simulated using \alpgen~\cite{bib:alpgen} with hadronic showering performed by \pythia~\cite{bib:pythia}.  
For signal and background modeling we use the \cteq\ set of parton distribution functions (PDFs)~\cite{bib:CTEQ}. 
The normalization of the \wpj\ contribution is a free parameter in the fitting procedure described below. 
 Events with multiple jets can also mimic  \ttbar\ signal when
 a particle from one of the jets is misidentified as an isolated lepton. 
The normalization of this multijet background is extrapolated from a  control sample enriched in this process
using the probability for a jet to satisfy the lepton-quality requirements~\cite{bib:matrix_method}. 
For the other backgrounds, $Z$+jets events are simulated with \alpgen, 
diboson events are simulated with \pythia, and events from single-top-quark production
are simulated with \comphep~\cite{bib:comphep}. The normalizations for the last three background
processes are taken from NLO calculations~\cite{bib:mcfm}.
In all cases, event generation is 
followed by the  \geant\-/based \DZ\ detector simulation~\cite{bib:geant}.  
To model energy depositions from noise and additional \ppbar\ collisions within the same bunch crossing, 
simulated events are overlaid with data from random \ppbar\ crossings.
All simulated events are reconstructed using the same code as for the reconstruction of the collider data. 

%
%
\section{ Reconstruction of the event kinematics}
\label{sec:reco}
To measure the \ttbar\ forward--backward asymmetry and its dependence on the invariant mass of the \ttbar\ 
system we need to determine the four-vectors 
of top quark and antiquark, which is done by summing the four-vectors of their decay products in the 
$\ttbar\to W^+b W^-\bbar$; $W^+ \to l^+ \nu_l$;  $W^- \to q\qbar'$ (and charge conjugates) decay chain.
There are four final state quarks in this decay chain, 
while we select events that contain one isolated lepton and at least three jets. 
When an event contains at least four jets we assume that the four jets with the largest \pt\ originate from the 
quarks from \ttbar\ decay. 
If an event contains only three jets one of the jets from \ttbar\ decay is missing. 
In either case all possible assignments of three or four jets to the final state quarks are used, 
with the likelihood of each assignment evaluated by the reconstruction algorithms described below.
 
For \lpgefj\ events, the \ttbar\ system is fully reconstructed using a kinematic fitting algorithm.
Previous \DZ\ top quark analyses used the algorithm of Ref.~\cite{bib:hitfit}. 
In this \this\ a new algorithm is employed, which utilizes an analytic solution for the neutrino momentum 
using the constraints on the $W$-boson ($M_W$) and top-quark masses ($m_t$)~\cite{bib:analytical}. 
The likelihood term for each jet-to-quark assignment  accounts for the differences between
the observed jet energy and the energy scaled to satisfy the constraints on $M_W$ and $m_t$.
The jet energy resolution and the probability for a jet 
to be reconstructed (see ``Type III'' transfer function in Ref.~\cite{bib:transfer_funcs}) are taken into account.
The $b$-tagging observables \mvab\ are also used to evaluate the likelihood of each assignment.

For \lptj\ events, a partial reconstruction algorithm of the \ttbar\ decay chain is employed~\cite{bib:NIM_3j}.
With one jet entirely lost, no significant improvement is expected from scaling the
four-vectors of the remaining objects as is done by the kinematic fitting algorithm in \lpgefj\ events, so the partial reconstruction algorithm does not attempt
to modify the kinematics of the  observed objects. As only the transverse components of the
neutrino momentum are measured in \met, the longitudinal component is
calculated using a quadratic equation which results from imposing the $M_W$ constraint on the $W\to l\nu$ decay products.  
The two-fold ambiguity is resolved by choosing the solution that minimizes the difference between the known $m_t$ and
the invariant mass of the objects assigned to the leptonic top quark decay, $m_l$. 
This algorithm thus assumes that the jet associated with the $b$ quark from the leptonically decaying top quark is detected.
This assumption holds for 80\% of the \ttbar\ events.
The lost jet is assumed to be associated with either a light quark or a $b$ quark from the hadronic top-quark decay chain.
In the majority of  cases (74\%) this jet is lost due to its low energy, 
so this loss has little effect on the kinematics of the hadronically decaying top quark. 
The lost jet is neglected in the partial reconstruction algorithm.
The sum of the four-vectors of the two jets assigned to the products
of the hadronically decaying top quark serves as a proxy for the four-vector of the hadronically decaying top quark with the invariant mass $m_p$.
Even though $m_p$ is not expected to be equal to $m_t$,  
the distribution in this variable is different for combinations correctly associated with the hadronically decaying top quark and 
 combinations that include a $b$ jet from the leptonically decaying top quark. 
In each event we consider the following nine observables:  the \mvab\ for each of the three jets, 
the three possible $m_l$, corresponding to the three possible lepton-neutrino-jet combinations,
and  the three possible $m_p$.
The likelihood of each of the three possible jet-to-quark assignment is calculated by evaluating the consistency of the nine observables 
with the distributions corresponding to the hypothesized assignment.
In particular, the jet hypothesized to be associated 
with a $b$ quark should have a value of \mvab\ consistent with the one expected for $b$ jets, 
while for a jet hypothesized to originate from
a $W$ boson decay \mvab\ should be consistent with the distribution expected for such jets. 
The values of  $m_l$ and $m_p$ for the jet combinations that correspond to the hypothesized assignment 
should be consistent with the distributions expected for correctly assigned jets, 
while the values of  $m_l$ and $m_p$ for the other jet combinations should agree with the distributions 
expected for wrong assignments. 
 When calculating \mtt, we compensate for the effect of the lost jet by applying an $m_p$-dependent scaling  
to  the four-vector of the hadronically decaying top quark. 

Unlike the \afb\ measurement in Ref.~\cite{bib:ourPRD}, where only the jet-to-quark assignment with the lowest \chisq\ was used, we reconstruct
 \dy\  by averaging its values over
all possible assignments, weighted by their likelihoods evaluated as described above for \lpgefj\ and \lptj\ events.  
The same approach is used to reconstruct \mtt\ in the \lptj\ channel.
For \lpgefj\ events, \mtt\ is reconstructed using the outputs of three reconstruction algorithms:
the new kinematic fit algorithm, the kinematic fit algorithm of Ref.~\cite{bib:hitfit}, 
and a simple reconstruction algorithm~\cite{bib:ttres} that evaluates the kinematics of the leptonically decaying $W$ boson from  
the lepton and the neutrino by imposing the $M_W$ constraint 
and calculates \mtt\ by adding the four most energetic jets without imposing the $m_t$ constraint.
The likelihood values calculated by the algorithms give indications on how well the kinematics of a particular event match the 
assumptions made by a given algorithm. 
In particular, for high \mtt\ there is a higher probability that two final state quarks are associated with the same jet. 
Such a jet is likely to be the most energetic jet in the event and have a large mass. 
The simple reconstruction algorithm, which does not assume a specific jet-to-quark assignment, performs best for such events.  
We use a multivariate regression~\cite{TMVA2007} to combine the partially correlated
\mtt\ values and the likelihoods produced by the three algorithms with 
supplementary observables such as the mass of the leading jet to estimate \mtt.
This combined \mtt\ reconstruction outperforms the individual algorithms in all \mtt\ ranges.

For the asymmetry measurement the performance of a \ttbar\ reconstruction algorithm can be characterized by the 
probability \pcor\ to correctly reconstruct the sign of \dy.
For the algorithm employed in this analysis for  \lpgefj\  events $\pcor=0.775$, 
compared to $\pcor=0.756$ for the algorithm of Ref.~\cite{bib:hitfit}.  
The partial reconstruction algorithm achieves $\pcor=0.745$ for \lptj\ events. 
The dependence of  \pcor\ on the \prodhlvl\ \absdy\ is shown in Fig.~\ref{fig:dil_comp_p}  for these three algorithms. 
The  high values of \pcor\ achieved by the partial reconstruction algorithm, which are almost as high as \pcor\ for \lpgefj\ events, can be understood from the following consideration. 
All four leading  jets are associated with the  quarks from the \ttbar\ decay in only 55\% of  the \lpgefj\ events. 
For the other 45\% of the events  one of the jets originates from initial or final state radiation, 
which can lead to badly misreconstructed \ttbar\ four-vectors. 
Only 4\% of the  \lptj\ events contain a jet that does not originate from the four quarks of the \ttbar\ decay. 
Thus, even though some information is lost with the unreconstructed jet, no wrong information is added, leading
to a low probability to misreconstruct the sign of \dy.  

\begin{figure}[htbp]
\begin{center}
\includegraphics[width=0.95\linewidth]{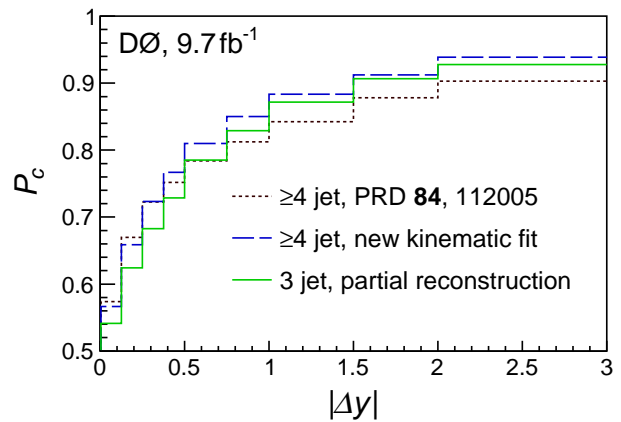}
\end{center}
\vspace{-0.4cm}
\caption{
(Color online). The probability to correctly reconstruct the sign of \dy\ as a function of the \prodhlvl\ \absdy\ 
for the algorithm of Ref.~\cite{bib:hitfit} used to measure the \afb\ in Ref.~\cite{bib:ourPRD} 
and the algorithms used to reconstruct \lpgefj\ events and to partially reconstruct \lptj\ events in this \this.
}
\label{fig:dil_comp_p}
\end{figure}

%
%
\section{SM predictions for \afb}
\label{sec:Preds}

The differential cross section for \ttbar\ production is 
available only at order $\alpha_s^3$, where $\alpha_s$ is the strong coupling constant. Since in the SM the \ttbar\ asymmetry only appears 
at this order, no full higher order prediction for \afb\ exists yet. 
The relative uncertainty  on the $\alpha_s^3$ calculation of the asymmetry due to higher order corrections is
evaluated to be as large as $\approx25\%$~\cite{bib:Bern}.

Recently the order $\alpha_s^4$ calculation for  the total cross section of \ttbar\
production~\cite{bib:Mitov} was made available, but the asymmetry was not computed at this order. 
Several papers report calculations of the leading
corrections  to the asymmetry with the predicted \afb\ values ranging from $5.0\%$ in \mcatnlo\ to $8.8\%$
once the electroweak corrections~\cite{bib:EWcorr} and 
resummations of particular regions of phase space~\cite{bib:resummed_afb} are taken into  account. 
The dominant uncertainty on these predictions is from the renormalization and factorization scales,
and is evaluated to be as high as $2.0\%$ (absolute)~\cite{bib:MCFMttbar,bib:Bern}.
The authors of Ref.~\cite{bib:PMC_scale} obtain a value of $\afb=12.7\%$ by choosing a normalization scale 
that  arguably stabilizes the perturbative expansion yet differs significantly from the scales commonly 
used in top quark physics calculations.
Some authors suggest that the corrections from interactions between the top quark decay products and the proton remnants 
should also be taken into account when calculating \afb~\cite{bib:rescat}.
Given this variety of predictions, we choose to compare our data to the well defined  \mcatnlo\ simulation.

At order $\alpha_s^3$,  the QCD contributions to the asymmetry in  \ttbar\ production can be divided into two classes
up to divergences that cancel between these two classes~\cite{bib:KnR}. The first class, which  contributes to negative asymmetry,
is a result of  interference  between the terms that contain gluon radiation in the initial or final states, which 
may result in an extra jet in the event and typically leads to a higher transverse momentum of the \ttbar\ system.  
The second class, which contributes to positive asymmetry, is from interference between the Born term ($\alpha_s^2$)
and the term described by a box diagram ($\alpha_s^4$).  The overall asymmetry is positive and depends on the jet multiplicity. 
Selection criteria that give preference to events with higher jet multiplicity favor the first class of events and further
lower the overall expected asymmetry, while a higher asymmetry is expected for events with lower jet multiplicity. 
Consequently,  forward events tend to have fewer jets than events in the backward category. 
Similarly, since a $b$-tagged jet is less likely to originate from initial or final state radiation, samples with  
a larger number of  $b$ tags tend to have higher values of \afb.

\mcatnlo\  predicts an overall asymmetry in \ttbar\ production before selection of $\left(5.01\pm 0.03\right)$\%. Here, and in the following sections, 
the quoted uncertainties on the predictions are from the finite size of the simulated samples unless otherwise stated. 
Table~\ref{tab:preds} lists the \mcatnlo\ predictions for  \ttbar\ events after the selection criteria are applied.

All previous measurements of \afb\ in the \lpj\ channel selected \ttbar\ events that had at least four jets in the final state. 
As is apparent from Table~\ref{tab:preds}, restricting the selection to only \lpgefj\ events lowers the
\prodhlvl\ asymmetry.
Including events with three jets reduces this selection bias.

\begin{table}[htbp]
\caption{ Asymmetries predicted by \mcatnlo\ for  \ttbar\ events that pass the analysis selection criteria. Statistical uncertainties only.
 }
\begin{ruledtabular} 
\begin{tabular}{lc@{\extracolsep{2ex}}c}
                &  \multihead{2}{\afb, \%}\\
\head{\multirow{2}{*}{Channel}} & \head{Production  } & \head{Reconstruction}\\
                                & \head{level} & \head{level} \\
\hline
\getj, \geob & $4.7\pm0.1$  & $3.9\pm0.1$ \\
\hline
\threej, \oneb &  $6.6\pm0.2$  &  $4.7\pm0.3$ \\
\threej, \getb &  $7.3\pm0.2$  &  $5.6\pm0.2$\\
\hline
\gefj, \oneb &  $1.4\pm0.2$  &  $1.9\pm0.2$\\
\gefj, \getb &  $3.2\pm0.1$  &  $3.3\pm0.2$\\
\end{tabular}
\end{ruledtabular}
\label{tab:preds}
\end{table}

Asymmetries after reconstruction are presented in the last column of Table~\ref{tab:preds}.
Finite resolution in \dy\  results in roughly $20\%$ of the forward events being misreconstructed as backward, and vice versa. 
Since there are more forward events,  \dy\ smearing leads to an overall lowering of the 
reconstructed asymmetries.
At the same time,  forward \ttbar\ events, which  
tend to have fewer jets, have a lower probability  to be misreconstructed, resulting in fewer migrations into
the backward  category, and an upward shift in the  reconstructed asymmetry. 
This bias is  most apparent in the \lpgefj, one-$b$-tag channel, where the lowest asymmetry
is predicted.

%
%
\section{Sample Composition and \Recohlvl\ \afb}
\label{sec:bg}

Reconstructed events are divided into six channels by the number of jets and $b$ tags: \lptj\ and \lpgefj\ with 0, 1, and $\ge$2 $b$ tags each. 
The \lptj\ zero-$b$-tag channel is used only for the background asymmetry calibration, 
and not for the \ttbar\ asymmetry measurement. 
The \lpgefj\ zero-$b$-tag channel is used only for determining the sample composition and the \recohlvl\
\afb, and is not used for measuring the \prodhlvl\ asymmetry. 

Several well-modeled variables that have different distributions for signal and background processes, 
and that have minimal correlations between each other and with \dy\ and \mtt, 
are combined into kinematic discriminants bounded between 0 and 1~\cite{bib:our_afbl}.
For \lpgefj\ events a discriminant $D_4$ is built from the following input variables:
\begin{itemize}
\item \chisq\ -- the test statistic of the likeliest assignment from the kinematic fit. 
\item \lbpt\ -- the transverse momentum of the leading $b$-tagged jet, 
or when no jets are $b$ tagged, the \pt\ of the leading jet.
\item $\ktmin=\min\left(p_{T,a},p_{T,b}\right)\cdot\Delta{\cal R}_{ab}$, where
$\Delta{\cal R}_{ab}=\sqrt{\left(\eta_a-\eta_b\right)^2+\left(\phi_a-\phi_b\right)^2}$ 
is the angular distance\footnote{Here the pseudorapidity $\eta$ and the azimuthal angle $\phi$ are defined relative to the PV.} 
between the two closest jets, $a$ and $b$, and $p_{T,a}$ and $p_{T,b}$ are their transverse momenta.
\item \mjj, the invariant mass of the jets corresponding to the $W\to q \qbar'$ decay 
in the likeliest assignment from the kinematic fit, calculated using kinematic quantities before the fit.\end{itemize}
The variables \chisq\ and \mjj\ are based on the full \ttbar\ reconstruction using the kinematic fitting technique of Ref.~\cite{bib:hitfit}. 

For the \lptj\ events we construct a discriminant $D_3$ using a different set of input variables:
\begin{itemize}
\item $S$ --- the sphericity~\cite{bib:spher}, defined as 
  $S = \frac{3}{2} (\lambda_2 + \lambda_3)$, where  $\lambda_2$ and $\lambda_3$ are the two largest of the three 
  eigenvalues of the normalized quadratic momentum tensor $M$. The tensor $M$ is defined as
  \begin{equation}
    M_{ij} = \frac {\sum_o {p_i^o p_j^o}}  {\sum_o{|p^o|^2 }},
    \label{eq:tensor}
  \end{equation}
  where $p^o$ is the momentum vector of a reconstructed object $o$, and $i$ and $j$ run over the 
  three indices for the Cartesian coordinates. 
  The sum over objects includes the three selected jets and the selected charged lepton. 
\item \ptthree\ --- the transverse momentum of the third leading jet. 
\item \mjjmin\ --- the lowest of the invariant masses of two jets, out of the three possible
  jet pairings.
\item \lbpt, defined as for the \lpgefj\ channel, above. 
\item \djom, the difference in azimuthal angle between the leading jet and \met. 
 \end{itemize}

The discriminants for all channels are combined into a single discriminant $D_{c}$, so that 
for the \lptj\ events $D_c=\Ntag+D_3$, while for \lpgefj\ events $D_c=3+\Ntag+D_4$.
The variable \Ntag\ above is usually taken to be equal to the number of $b$-tagged jets
in the event, but for events with more than two $b$-tagged jets $\Ntag=2$ instead.
We fit the sum of the signal and background templates to the data distribution in the discriminant $D_{c}$ as shown in Fig.~\ref{fig:disc}.
This fit is identical to the fit for the sample composition in Ref.~\cite{bib:our_afbl}. 
The sample composition and its breakdown into individual channels are summarized in Table~\ref{tab:sample}. 
Background contributions other than \wpj\ and multijet production are labeled ``Other Bg'' in Table~\ref{tab:sample}.

\begin{figure*}[htbp]
\begin{center}
\includegraphics[width=0.75\linewidth]{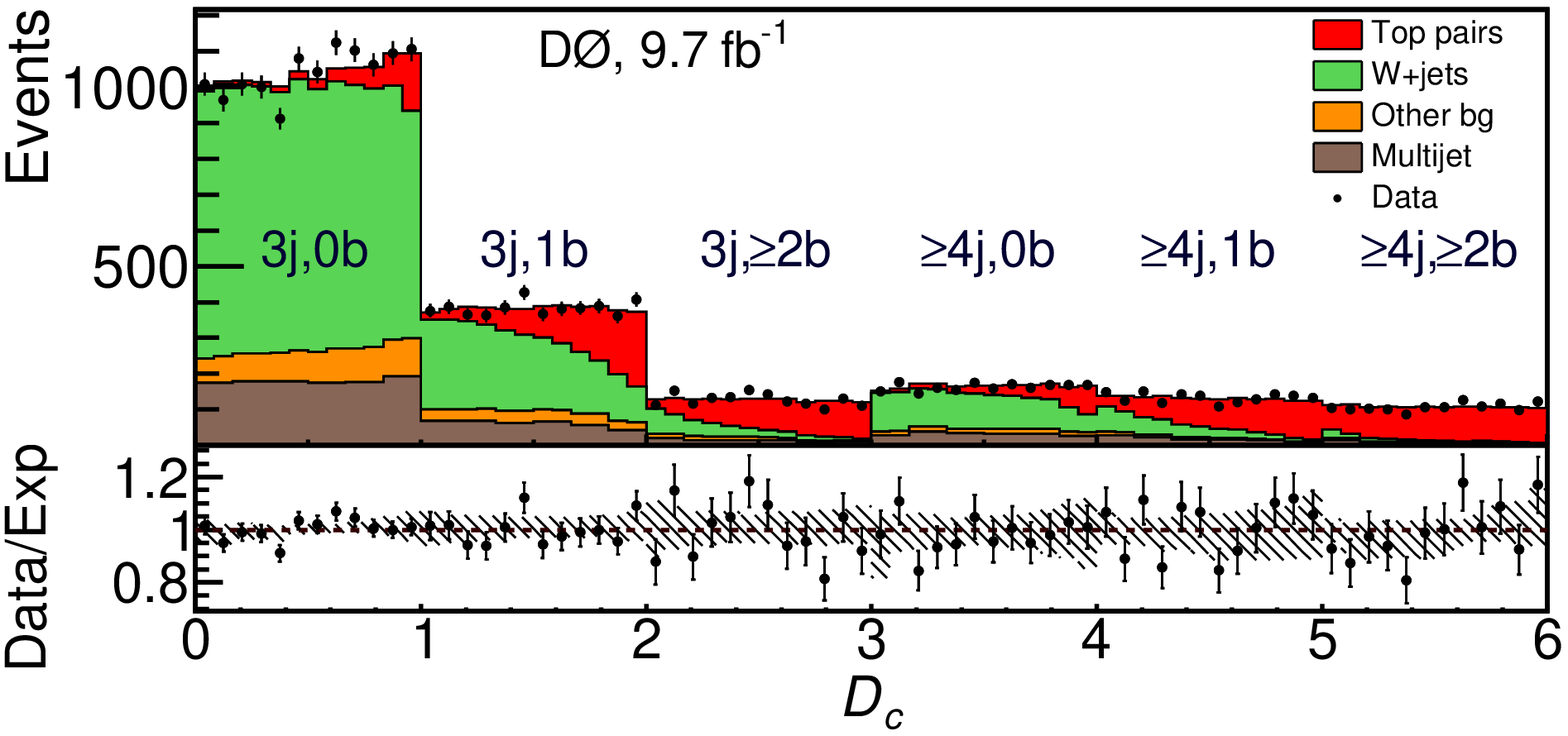}
\end{center}
\vspace{-0.5cm}
\caption{(Color online).
The combined discriminant  $D_{c}$.
The region $D_{c}<1$ is not used to determine the signal \afb.
\dvmLowerPanel\ Figure is from Ref.~\cite{bib:our_afbl}.
}
\label{fig:disc}
\end{figure*}

\begin{table}
\caption{
Estimated number of events from the fit of the data distribution in the discriminant $D_{c}$ to the sum of signal and background processes (see Fig.~\ref{fig:disc}).
The sum of the estimated number of signal and background events is constrained to be equal to that in data.
The ``Selected events'' column includes the \lptj\ events with at least one $b$ tag and all \lpgefj\ events. 
The statistical uncertainties from the fit are quoted.
We also present the event breakdown for the channels with at least one $b$ tag, which are used to determine the
\prodhlvl\ \afb. Table is from Ref.~\cite{bib:our_afbl}.}
\begin{ruledtabular} 
\begin{tabular}{lccccc}
  & \head{Selected}& \multihead{2}{\threej} & \multihead{2}{\gefj} \\
\head{Source} & \head{events} & \head{\oneb} & \head{\getb} & \head{\oneb} & \head{\getb}  \\
\hline
\wpj  & $4447 \pm 74$ & $2461$ & $352$ & $403$ & $79$\\
Multijet & $969 \pm 24$ & $449$ & $95$ & $127$ & $62$\\
Other Bg & 786 & $404$ & $112$ & $75$ & $32$\\
Signal & $4745\pm70$ & $1212$ & $1001$ & $983$ & $1166$\\
\hline
Sum & 10947 & 4526 & 1560 & 1588 & 1339\\
\hline
Data & 10947 & 4588 & 1527 & 1594 & 1281\\
\end{tabular}
\end{ruledtabular}
\label{tab:sample}
\end{table}

In the simulated \wpj\ background, the angular distribution of leptons from $W$-boson decay has a forward--backward asymmetry, which is in part tuned to Tevatron  data~\cite{bib:CDFWasym}.
Due to this asymmetry,  when these events are reconstructed according to the \ttbar\ hypothesis, 
there remains a residual  asymmetry of $\approx 5\%$ in the \dy\ distribution. 
To improve the modeling of this asymmetry, we apply a weight 
to each simulated \wpj\ event which depends on the product of the generated lepton charge  and 
its rapidity. These
weights are chosen so that the simulation best matches 
control data with three jets and zero $b$ tags as in Ref.~\cite{bib:our_afbl}.
The difference in the \dy\ distributions predicted by the simulation with and without the applied weights is treated as a source of systematic uncertainty due to background modeling. 
This uncertainty exceeds the uncertainty due to PDFs by about a factor of two. 
We rely on the simulation to predict
the variation of the asymmetry in \wpj\ events  with jet  and  $b$-tag multiplicities. 

\begin{figure}[htbp]
\begin{center}
  \includegraphics[width=0.48\linewidth]{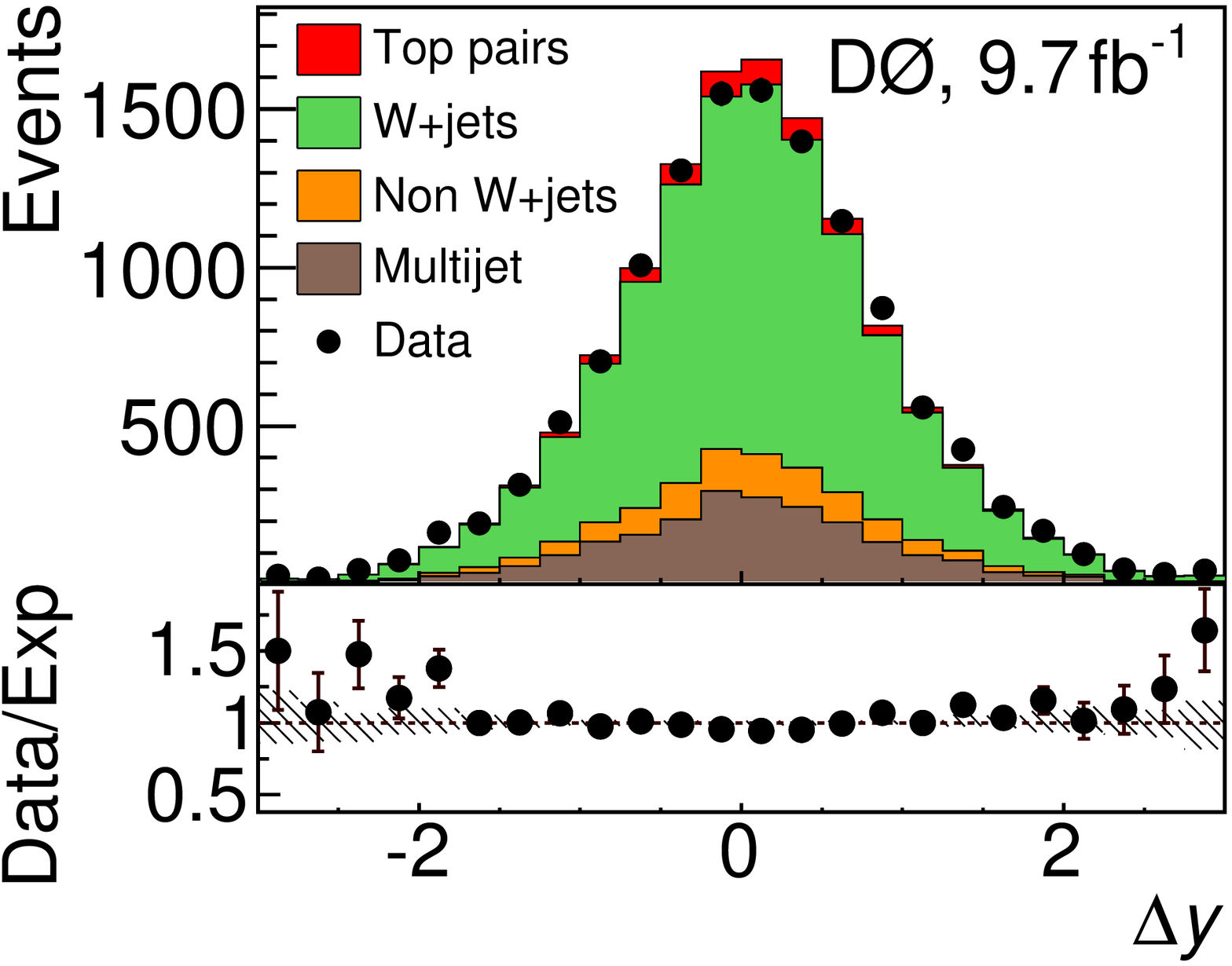}
  \includegraphics[width=0.48\linewidth]{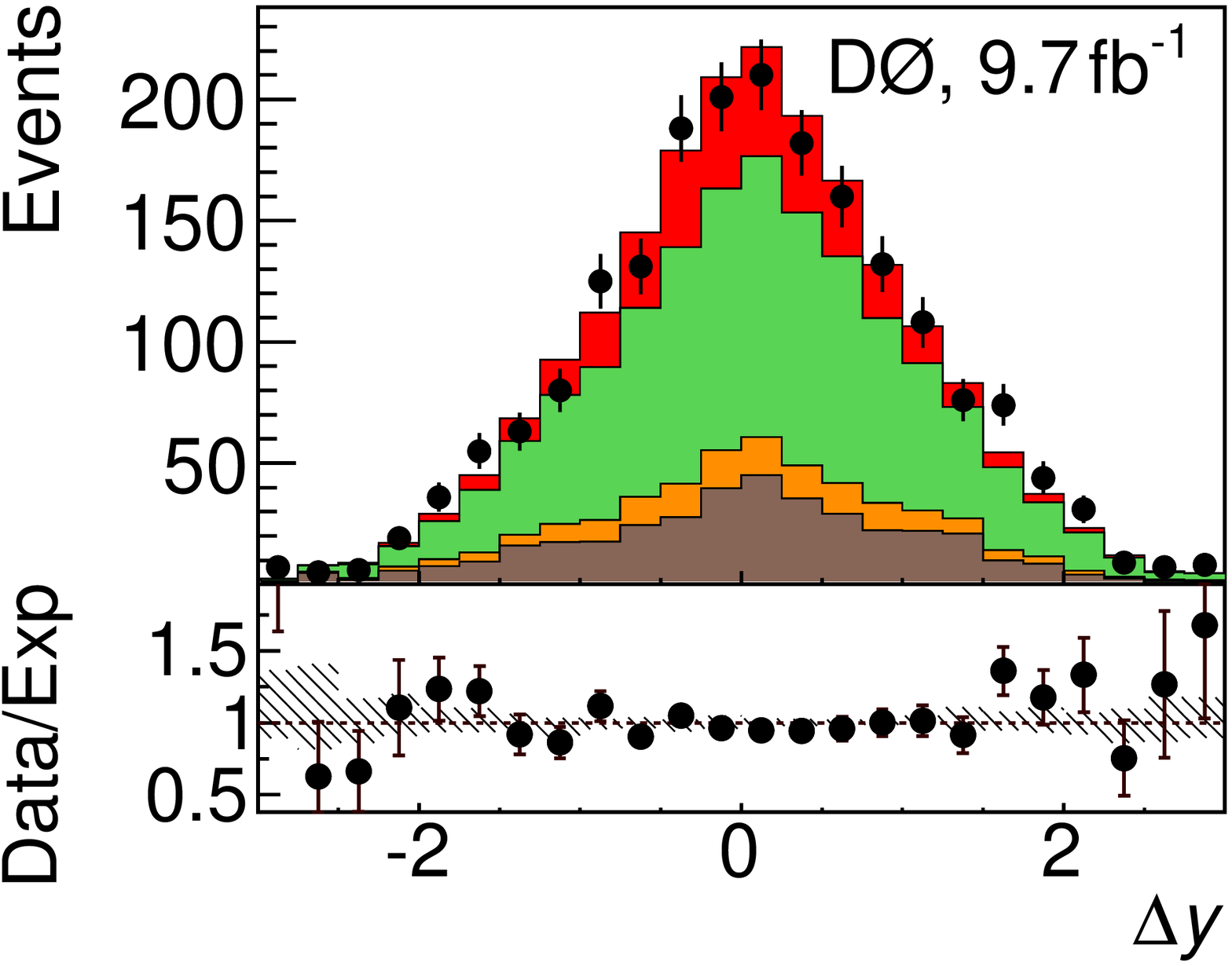}
  \includegraphics[width=0.48\linewidth]{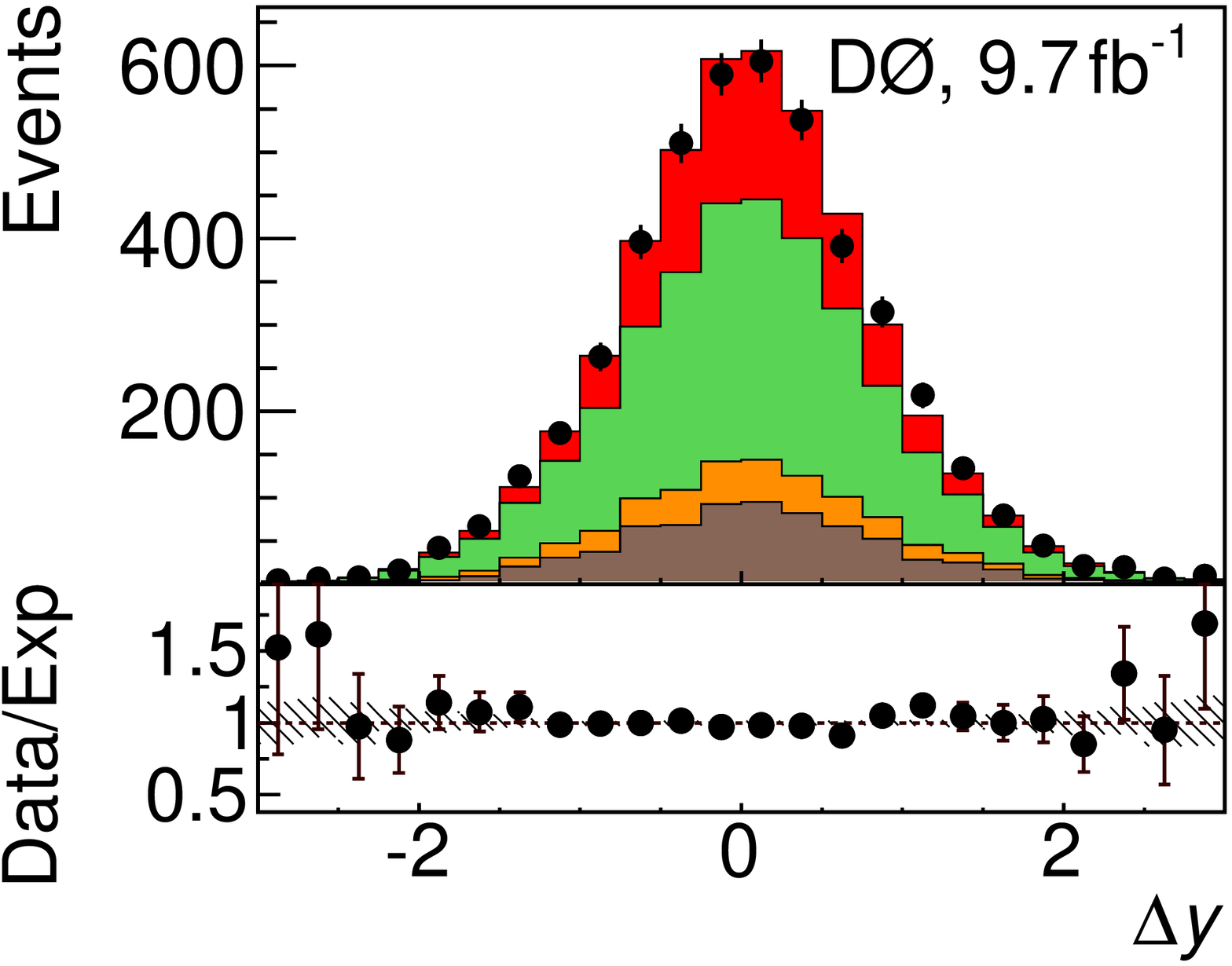}
  \includegraphics[width=0.48\linewidth]{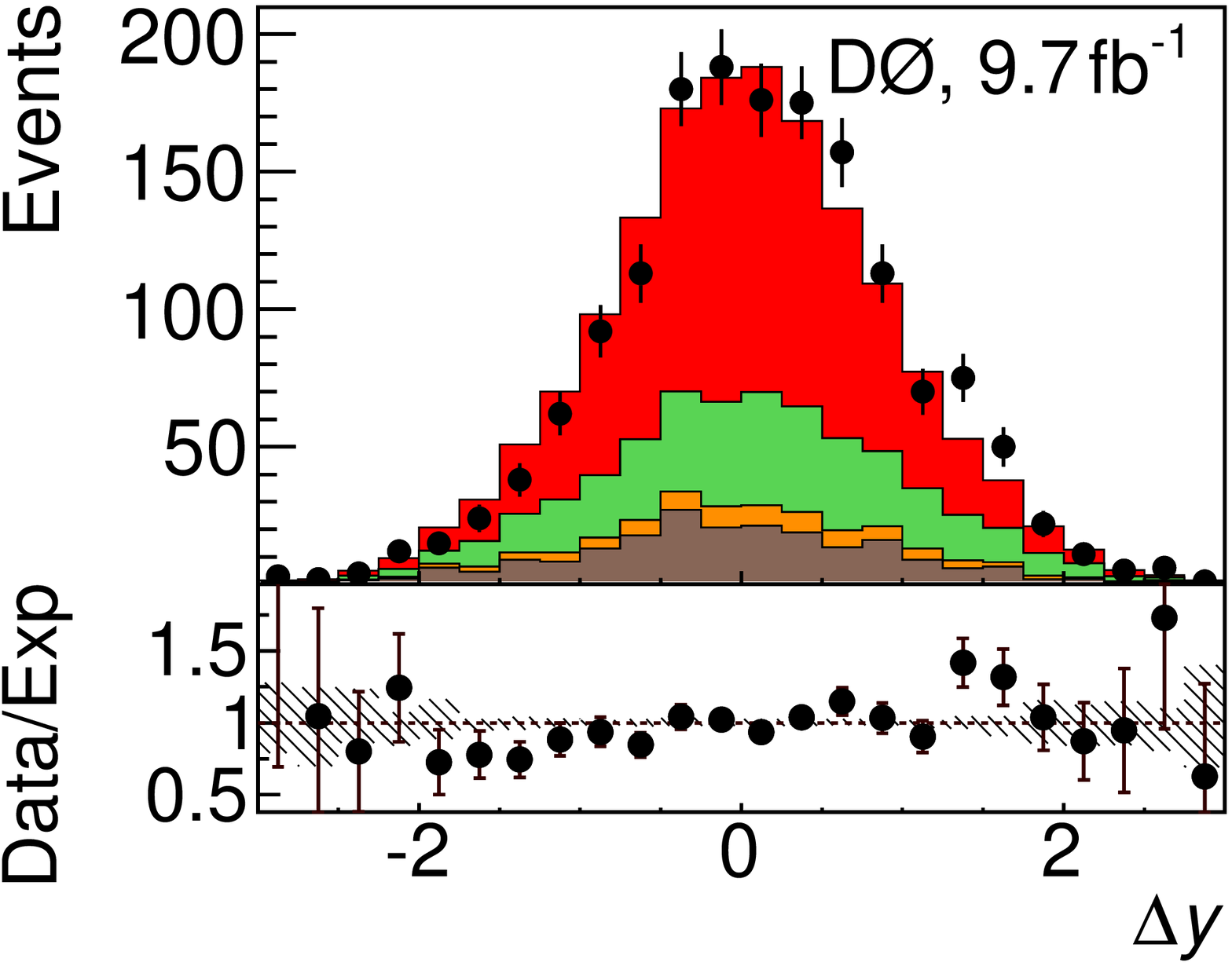}
  \includegraphics[width=0.48\linewidth]{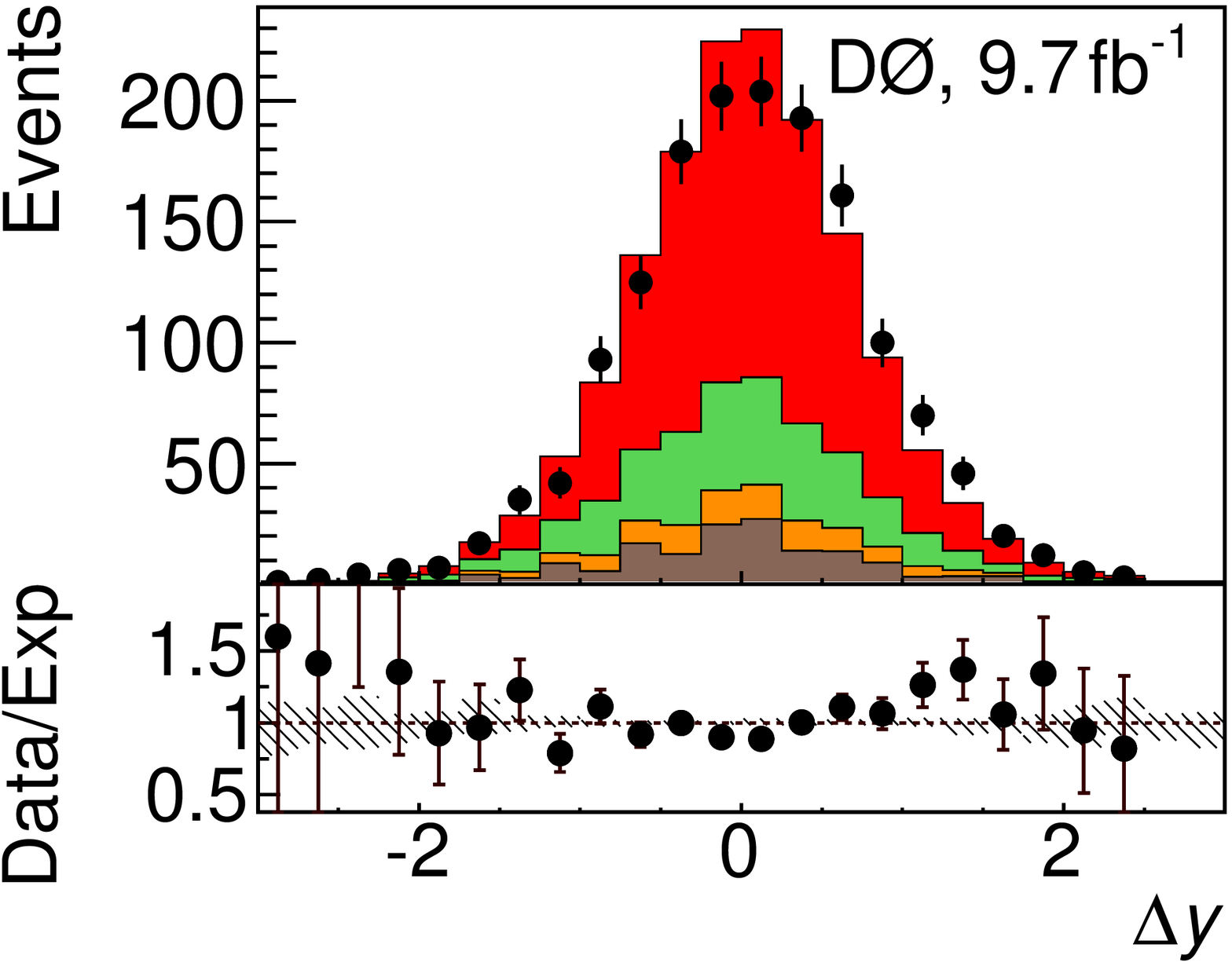}
  \includegraphics[width=0.48\linewidth]{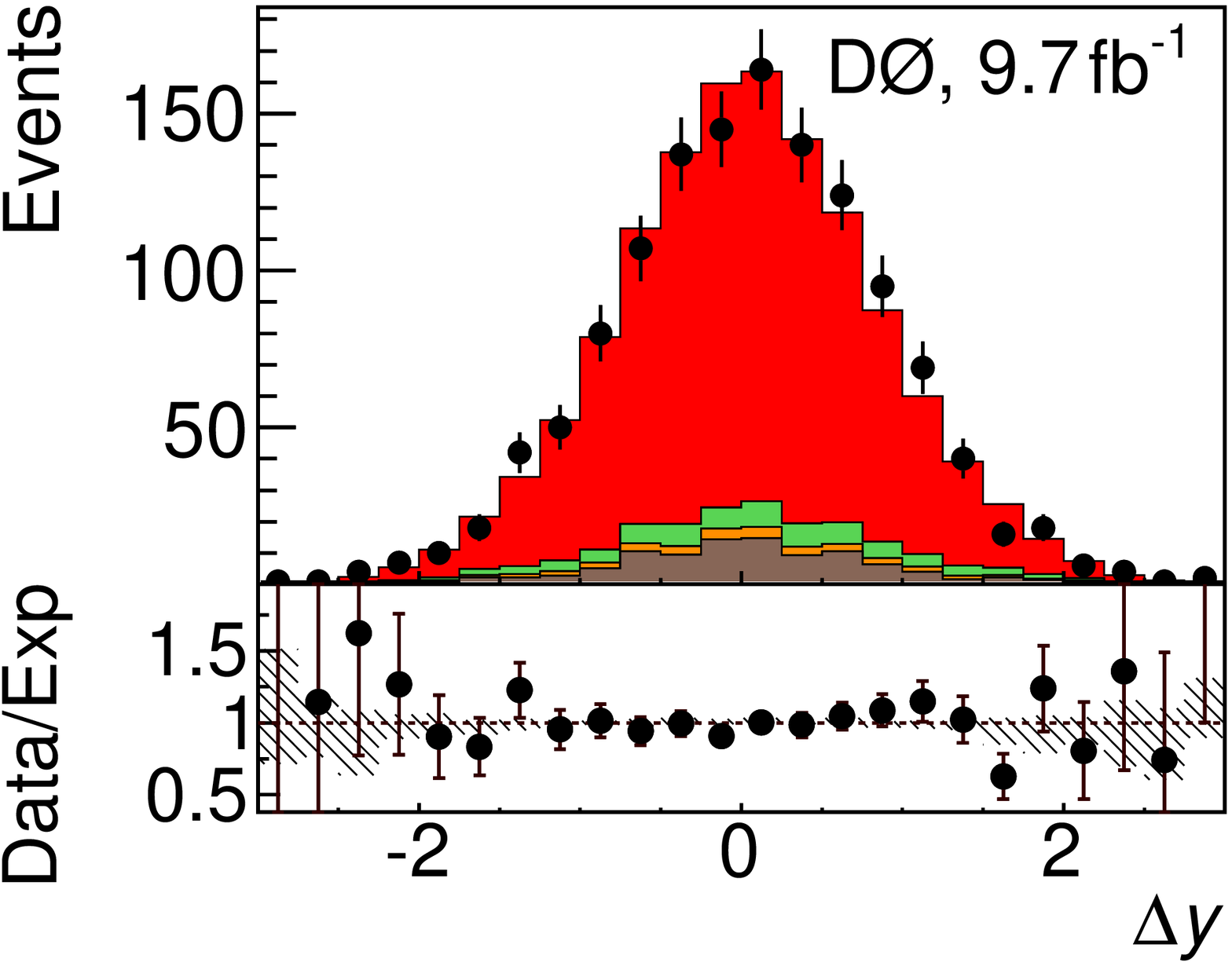}
\begin{picture}(132,5)(132,5)
\put (175,278){\subfloat[][]{\label{subfig:v3_a}}}
\put (295,278){\subfloat[][]{\label{subfig:v3_b}}}
\put (175,186){\subfloat[][]{\label{subfig:v3_c}}}
\put (295,186){\subfloat[][]{\label{subfig:v3_d}}}
\put (175, 96){\subfloat[][]{\label{subfig:v3_e}}}
\put (297, 96){\subfloat[][]{\label{subfig:v3_f}}}
\end{picture}
\end{center}
\vspace{-0.8cm}
\caption{(Color online).
  Reconstructed difference between the rapidities of the top and antitop quarks, \dy.
  \dvmLegendBoth\ Overflows are included in the edge bins.
  \dvmLowerPanels
}
\label{fig:dy_reco}
\end{figure}

The distributions of the reconstructed \dy\ are shown in Fig.~\ref{fig:dy_reco}. 
The \ttbar\ asymmetry at the reconstruction level is extracted using a fit to the distributions in the discriminant $D_c$ and  sign of \dy,
excluding the \lptj\ events with zero $b$ tags. 
This fitting procedure is identical to the procedure used in Ref.~\cite{bib:our_afbl}.
The inclusive asymmetry measured at the reconstruction level is $\left( 7.9 \pm 2.3\right)\%$. 
The results for individual channels are listed in Table~\ref{tab:recoasym}.

\begin{table}[htbp]
\caption[Reco.-level \afb s for selected \ttbar\ events for different subsamples.]
{\Recohlvl\ background\-/subtracted asymmetries for selected  events for different channels. 
The last row includes the channels listed in the rows above and the \lpgefj, zero-$b$-tag channel.
The first uncertainty is statistical, and the second one is systematic. 
Systematic uncertainties are discussed in Section~\ref{sec:syst}.
The prediction is based on the \mcatnlo\ simulation. }
\begin{ruledtabular} 
\begin{tabular}{lcc}
 & \multihead{2}{\afb, \%} \\
\head{Channel} & \head{Predicted} & \head{Measured} \\
\hline
\threej, \oneb & $4.7$ & $5.4 \pm 6.0 _{- 4.0}^{+3.3}$\tstrut\\
\threej, \getb & $5.6$ & \hphantom{$|$}$10.7 \pm 4.2 \pm 0.8$ \\
\gefj, \oneb   & $1.9$ & \hphantom{$|$}$11.0 \pm 4.4 \pm 0.8$ \\
\gefj, \getb   & $3.3$ & \hphantom{$|$}\noone$5.9 \pm 3.3\pm0.1$         \\
\hline
Combined       & $3.6$ & \hphantom{$|$}\noone$7.9 \pm 2.1 \pm 0.9$\tstrut  \\
\end{tabular}
\end{ruledtabular}
\label{tab:recoasym}
\end{table}

The distributions of the reconstructed invariant mass of the \ttbar\ system are shown in Fig.~\ref{fig:mtt_reco}. 
Since the \lptj\ and \lpgefj\ channels have different response (both mean and shape) 
for  \mtt, the dependence of \afb\ on \mtt\ at the reconstruction level
is difficult to interpret and is not presented here.
The measurement of \prodhlvl\ \afb\ and its dependence on \mtt\ is described in Section~\ref{sec:unfold}.

\begin{figure}[htbp]
\begin{center}
  \includegraphics[width=0.48\linewidth]{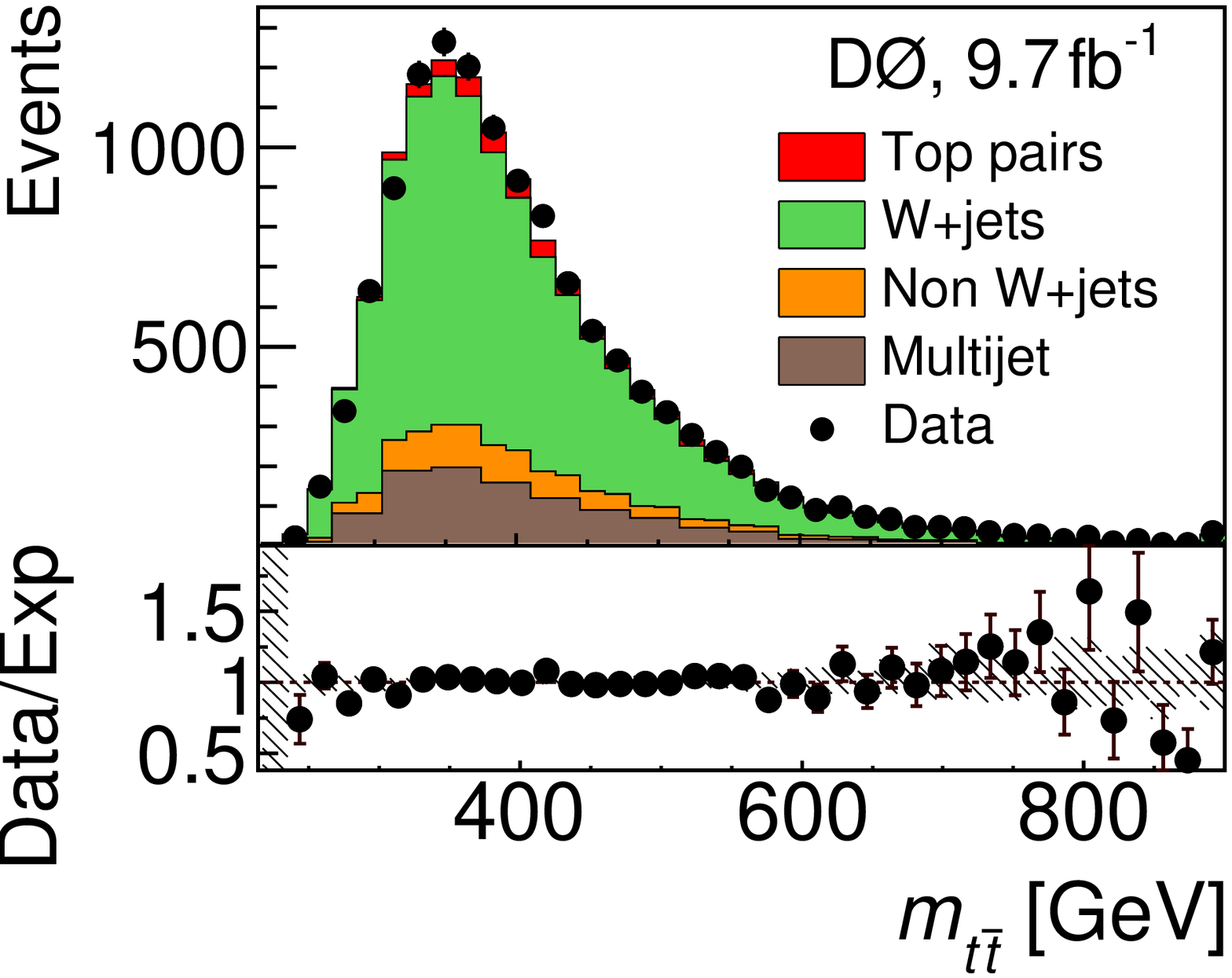}
  \includegraphics[width=0.48\linewidth]{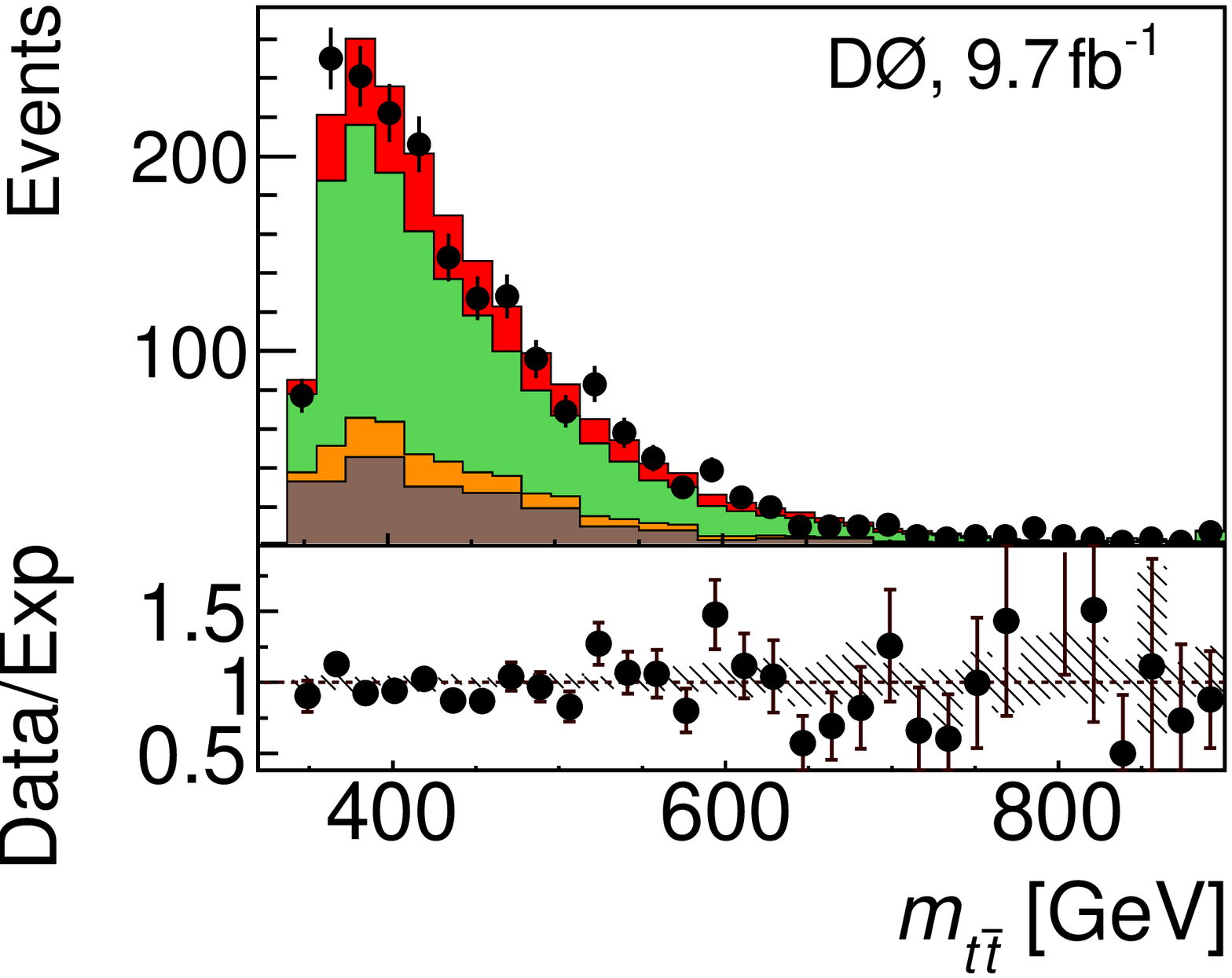}
  \includegraphics[width=0.48\linewidth]{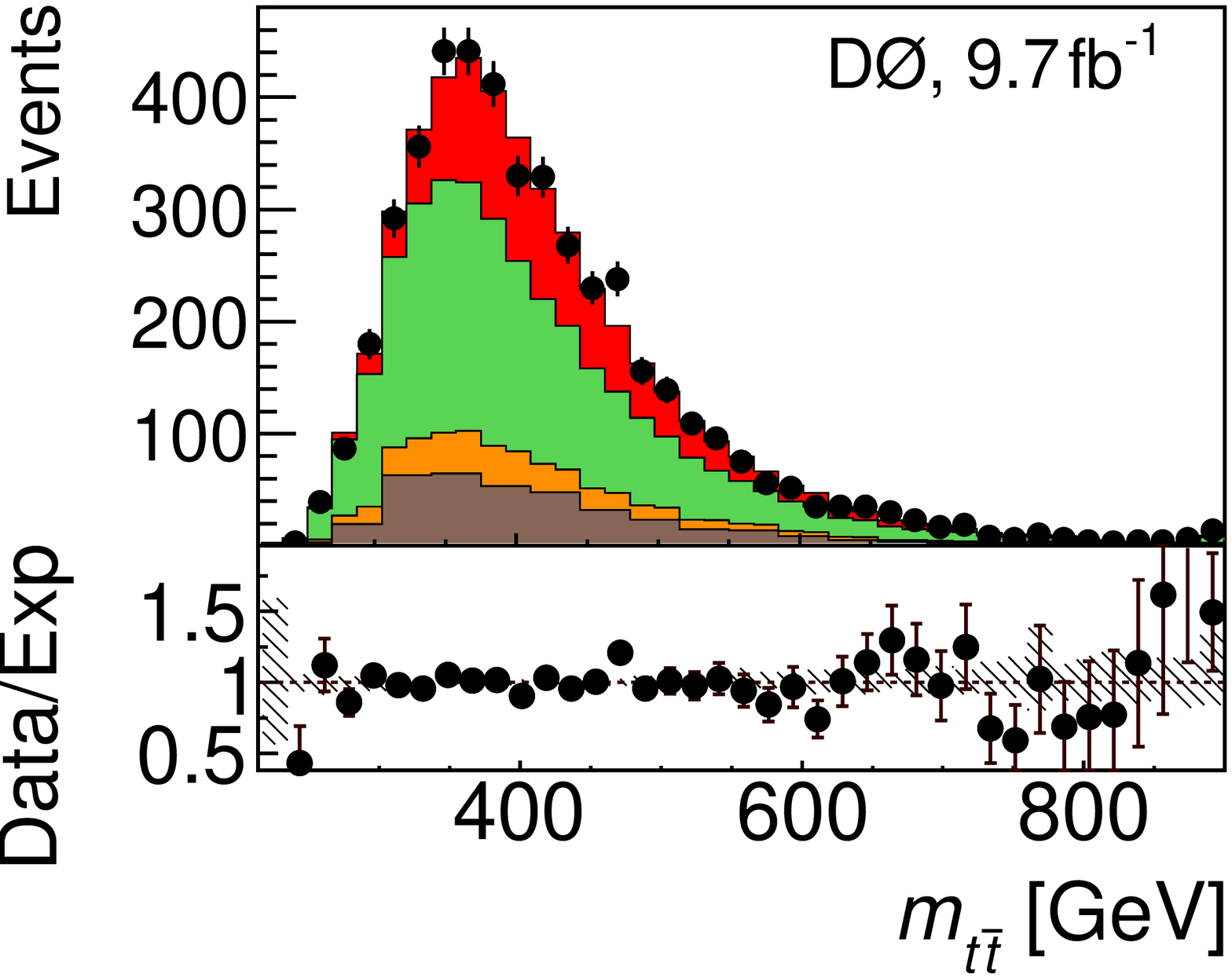}
  \includegraphics[width=0.48\linewidth]{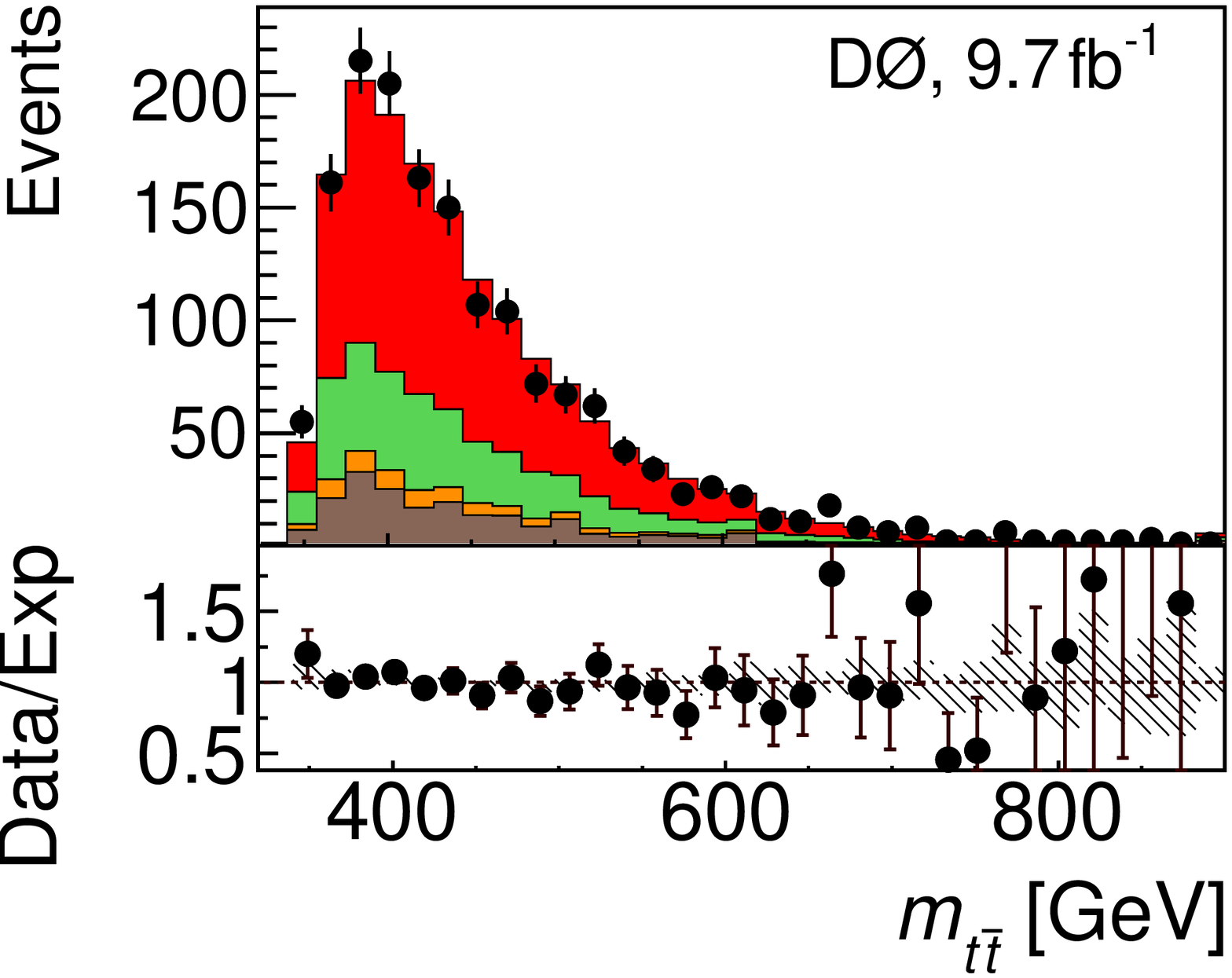}
  \includegraphics[width=0.48\linewidth]{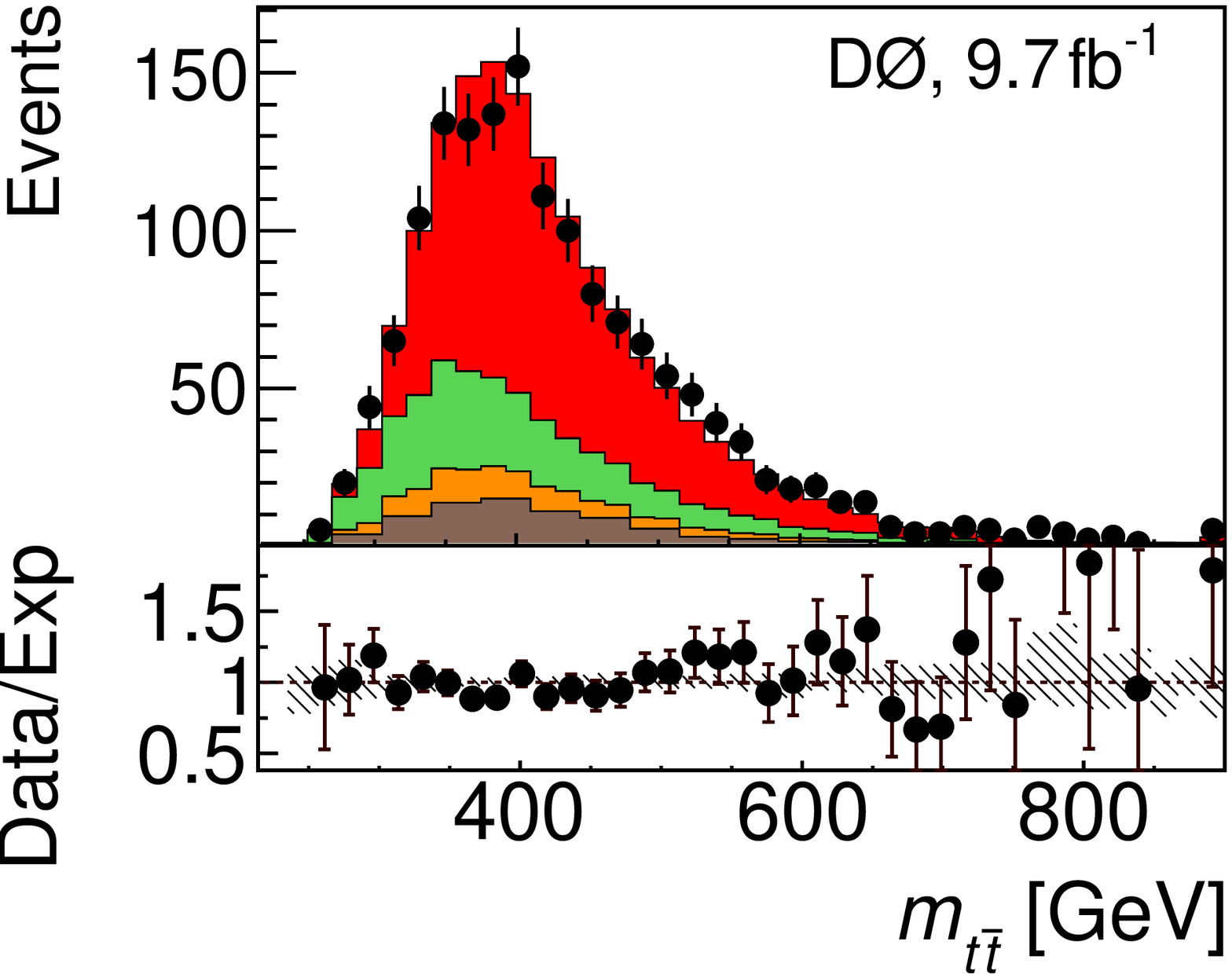}
  \includegraphics[width=0.48\linewidth]{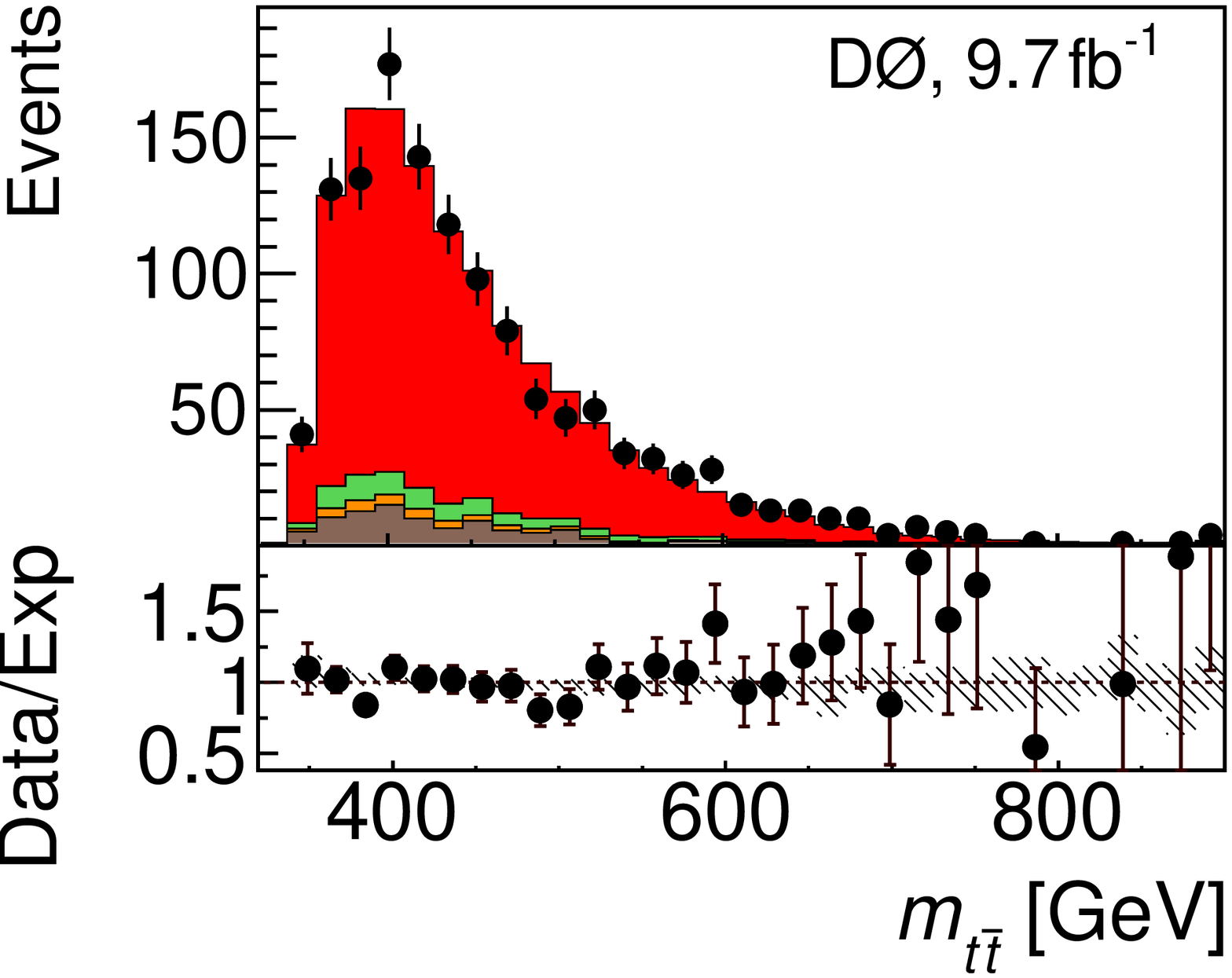}
\begin{picture}(132,5)(132,5)
\put (132,288){\subfloat[][]{\label{subfig:v4_a}}}
\put (293,278){\subfloat[][]{\label{subfig:v4_b}}}
\put (173,187){\subfloat[][]{\label{subfig:v4_c}}}
\put (293,187){\subfloat[][]{\label{subfig:v4_d}}}
\put (173, 96){\subfloat[][]{\label{subfig:v4_e}}}
\put (295, 96){\subfloat[][]{\label{subfig:v4_f}}}
\end{picture}
\end{center}
\vspace{-0.8cm}
\caption{(Color online).
  Reconstructed invariant mass of the top quark--antiquark pair, \mtt.
  \dvmLegendBoth\ \dvmLowerPanels
}
\label{fig:mtt_reco}
\end{figure}

We use the results of the sample composition study summarized in Table~\ref{tab:sample}
to normalize the distributions for the background processes in the sensitive variables 
(\dy, and also \mtt\ for the 2D measurement), which are subtracted from the distributions observed in data. 
To increase the signal purity of the data used in the fully corrected measurements, 
we unfold only events containing at least one $b$ tag. 
The background\-/subtracted distributions of the sensitive variables in the corresponding four channels are 
used as inputs to the unfolding procedure.

%
%
\section{Unfolding the asymmetry}
\label{sec:unfold}
The true or generated distribution of a certain variable (\dy\ for the inclusive measurement) is
shaped by acceptance and detector resolution, resulting in the observed distribution, 
which is also subject to statistical fluctuations.
The goal of the unfolding procedure is to find the best estimator for the true 
distribution given the \bsd\ and knowing detector acceptance and resolution from simulation.   
After finding the best estimator for the true distribution of \dy, we summarize it into the \prodhlvl\
 \afb\ using Eq.~\ref{eq:afb}.  This is the same general approach used in the previous measurement~\cite{bib:ourPRD}.
For this unfolding we use  \tunfold~\cite{bib:tunfold}, which we extend as discussed below.

Each distribution is presented as event counts in a binned histogram\footnote{Overflows are included in the edge bins.}, 
i.e., as a vector with a dimension equal to the number of bins. 
Given the vector of \prodhlvl\ \ttbar\ signal counts, $p$, and the vector of expected background counts, $b$,
the expected data counts in the $i$-th bin $\tilde{d}_i$ is given by
\begin{align}
\label{eq:exp_data}
 \tilde{d}_i = \migaccmat_{ij} p_j + b_i,\\
\migaccmat =  \migmat \accmat,
\end{align}
where \accmat\ is a diagonal acceptance matrix, 
whose $jj\text{-th}$ element is the probability for an event produced in the $j$-th bin to pass the selection criteria
and \migmat\ is the normalized migration matrix, 
whose $ij\text{-th}$ element is the probability for a selected event produced in the $j$-th bin to be observed in the $i$-th bin. 

Given the vector of observed counts, $d$, we can construct the vector of background\-/subtracted \recohlvl\ 
counts, $r=d-b$, with its covariance error matrix \Vrr.
The matrix \Vrr\ is constructed to account for the expected statistical uncertainties on data and background, 
in particular those due to the size of the multijet-enriched control sample. 
We then seek to find the vector $u$, 
which best estimates the vector of \prodhlvl\ counts $p$, by minimizing
\begin{align}
\label{eq:chisq}
\chisq & = (r-\migaccmat u)^T \Vrr^{-1} (r-\migaccmat u) + \tau^2 \left(\regmat u\right)^T \regmat u
\end{align}
for a given vector $r$, where $\tau$ is the regularization strength and
\regmat\ is the regularization matrix.
The first term of Eq.~\ref{eq:chisq} quantifies the consistency of $u$  with data, while
the second (regularization) term quantifies the  smoothness of $u$.

Without regularization, the unfolding procedure amounts to a 
minimization of the first term in Eq.~\ref{eq:chisq}. If the numbers of reconstruction and \prodhlvl\ 
bins are equal, the problem of minimization is solved by simply inverting the matrix: $u_{\text{unregularized}}=\migaccmat^{-1}r$.

Unregularized matrix inversion typically results in unphysical, rapidly varying distributions~\cite{bib:zechbook}.
Such distributions are disfavored in regularized unfolding by adding a second ``regularization'' term to the \chisq.
The regularization term in Eq.~\ref{eq:chisq} depends on the discrete second derivative of the binned distribution $u$. 
For constant bin widths, the regularization term is calculated
using a regularization matrix with the following structure~\cite{bib:tunfold}:
\begin{equation}
\label{eq:regM}
\regmat = \begin{pmatrix}
  0 & 0  & 0 & 0 &\cdots & 0 \\
  1 & -2 & 1 & 0 &\cdots & 0 \\
  0 & 1 & -2 & 1 & \cdots & 0 \\
  \vdots & & \ddots & \ddots & \ddots & \vdots \\
  0 & \cdots & 0 & 1 & -2 & 1 \\
  0 & \cdots & 0 & 0 &  0 & 0 \\
    \end{pmatrix}.
\end{equation}

For this analysis we modify the structure of \regmat\ to regularize based on 
the second derivative of the event density rather than the event counts, which allows
for the use of variable bin sizes. 
The regularization strength $\tau$ is chosen using both ensemble testing (described below)
and the L-curve technique~\cite{bib:tunfold} to balance the minimization of statistical fluctuations and
 bias. The difference between the two techniques is included in the evaluation of the systematic uncertainty due to the choice of the regularization strength. 

As in Ref.~\cite{bib:ourPRD}, 
the \prodhlvl\ \dy\ distribution is divided into 26 bins
and the \recohlvl\ \dy\ distribution is divided into 50 bins. 
Both have narrower bins near $\dy=0$, where the probability to misclassify forward
events as backward or vice versa changes rapidly, and 
wider bins at high \absdy, where statistics are low.

For the 2D measurement, we use six \mtt\ bins at the production level, with
edges at 0, 400, 450, 500, 550, 650 \GeV\ and $+\infty$.
The joint distribution of \dy\ and \mtt\ has a kinematic boundary at
$\absdy = \log\left(\left[1+\beta\right]/\left[1-\beta\right]\right)$, where 
$\beta  = \sqrt{1-\left(2m_t/\mtt\right)^2}$. 
A bin edge close to this boundary would result in a large difference in the event density between adjacent
bins, a feature that would be smoothed by a regularization procedure, thus biasing $u$.
To avoid such a bias, 
the \dy\ edges of the bins of the 2D measurement
are chosen to depend on \mtt\ as shown in Fig~\ref{fig:prod_bin}.

 The \recohlvl\ histograms have similar but finer bins along both the \dy\ and \mtt\ directions.
In the \lptj\ channels 13 \mtt\ bins are used to accurately describe migrations among the six 
\prodhlvl\ bins. 
The \mtt\ resolution in the \lpgefj\ channels allows for 14 \mtt\ bins.

\begin{figure}[htbp]
\begin{center}
\includegraphics[width=0.75\linewidth]{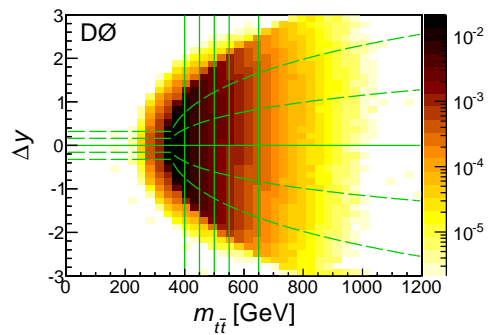}
\end{center}
\vspace{-0.6cm}
\caption[Prod.-level bins in the $\mtt,\dy$ plane.]{
(Color online). \Prodhlvl\ bins for the 2D measurement in the $(\mtt,\dy)$ plane, overlaid
on the distribution in these variables predicted from \mcatnlo. 
The shading reflects the predicted event density in arbitrary units.
The solid and dashed lines denote the \prodhlvl\ bins.
The solid lines show bins that are used for the final result.
}
\label{fig:prod_bin}
\end{figure}

We simultaneously unfold to the production level the four channels that contain at least one $b$ tag.
The difference in purity among channels is accounted for in the definition of the covariance error matrix \Vrr.
The unfolding technique is calibrated, and the statistical and systematical uncertainties are determined using the
results of ensemble tests. 
Each ensemble comprises simulated PDs that we build according to \mcatnlo, \alpgen~\cite{bib:alpgen} 
or \madgraph~\cite{bib:madgraph} SM predictions, or according to toy models with different asymmetries. 
The PDs are created from  the expected bin counts $\tilde{d}_i$ calculated using Eq.~\ref{eq:exp_data} by adding
Poisson (statistical) and Gaussian (systematic) fluctuations, with the Gaussian width taken as one standard deviation for the corresponding  systematic uncertainty.

In the toy models the input  distribution $P(\dy)$ has the form:
\begin{equation}
\label{eq:reshape}
P(\dy)=G\left(\dy;\mu,w\sigma_0\right) \left(1+a\erf\left(\dy/\delta\right)\right),
\end{equation}
where $a$ and $\delta$ are shaping parameters,
$G$ is a Gaussian distribution with mean $\mu$ and width $w\sigma_0$, 
$\sigma_0$ the width predicted by \mcatnlo, and $w$ a scaling parameter.
The shape of the \dy\ distribution and the input asymmetry are varied using
the parameters $\mu$, $a$, $\delta$, and $w$. 
In addition, we produce ensembles with the signal taken from simulated samples of \ttbar\ production mediated by axigluons, 
 hypothetical massive particles that arise in extensions of the SM that suggest different 
strong couplings for left and right-handed quarks~\cite{bib:axigluon}. 
The input asymmetry in the models used for calibration ranges from $-30\%$ to $+30\%$, while the axigluon masses 
are varied from $0.2$ to $2\TeV$. 

\begin{figure}[htbp]
\begin{center}
\includegraphics[width=0.75\linewidth]{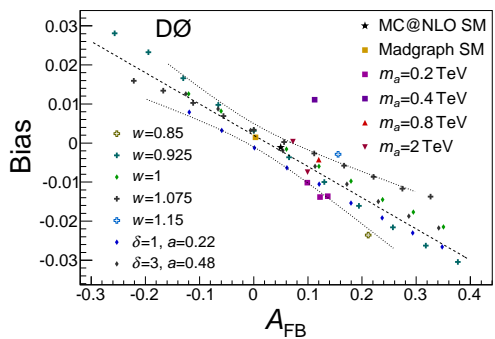}\\
\end{center}
\vspace{-0.6cm}
\caption[Calibration of inclusive \afb.]{(Color online).
The bias as a function of the input \afb.
Axigluon scenarios are indicated in the legend by the mass of the axigluon, $m_a$.
The toy models are labeled by parameters $w$, $a$, and $\delta$ of Eq.~\ref{eq:reshape}. 
Unless stated otherwise $w=1$, $a=0$, and $\delta=1$.
For each set of  $a$, $\delta$, and $w$, the value of $\mu$ is varied to produce different input asymmetries.
The dashed line indicates the calibration applied to the inclusive measurement
and the dotted lines indicate the assigned calibration uncertainties. The point significantly outside of the dotted lines corresponds to 
an axigluon mass of 0.4 \TeV\ and is discussed in the text. 
}
\label{fig:cal_example}
\end{figure}
The bias, which is the average difference between the unfolded and input \afb\ values,  
is shown in Fig.~\ref{fig:cal_example} as a function of the input \afb. 
Based on this study we derive a correction (calibration) that is applied to the result to eliminate the expected bias. 
The majority of the tested models are contained within the systematic uncertainty assigned to this calibration, 
which is shown by the dotted lines in Fig.~\ref{fig:cal_example}.
The one point that is significantly outside of the boundaries of this region corresponds to \ttbar\ production mediated by an axigluon with a mass of $0.4\TeV$.  
This particular model exhibits a significant change in \afb\ 
on the \mtt\ scale smaller than the bin width (here, 50\GeV), 
thus breaking the assumption of a smooth underlying distribution, leading to biased results. 
The unfolding of models with such rapidly changing \afb\ will, in general,
be biased in all regularized unfolding procedures, and we choose not 
to assign a systematic uncertainty that covers this specific class of models.

The unfolded \dy\ distribution is presented in eight bins, with 
each bin calibrated for the expected bias observed in the \mcatnlo\-/simulated PDs.
The \afb\ value in each \absdy\ range, \aady, is calibrated using the same procedure as for the inclusive \afb.
Since no systematic correlation is found between the \afb\ biases in different \absdy, as well as \mtt\ bins, 
they are calibrated individually.

%
%
\section {Systematic uncertainties}
\label{sec:syst}

The systematic uncertainties on the reconstruction and \prodhlvl\ \afb\ are summarized in Table~\ref{tab:sysAfb} in several  categories, which
are detailed below. To evaluate the systematic uncertainty on the \recohlvl\ \afb, we vary the modeling according to the estimated uncertainty in the relevant parameter of the model
and propagate the effect to the result. 
The systematic uncertainties on the \prodhlvl\ \afb\ are evaluated by including the effects of systematic variations on the simulated
background-subtracted PD into the ensemble tests. 
To find the expected uncertainty due to each category we use dedicated ensembles 
generated without statistical fluctuations and with only the relevant systematic effects.
The total uncertainties on the \prodhlvl\ \afb\ are taken from ensembles built including both statistical fluctuations 
and systematic effects (see Sec.~\ref{sec:unfold}).

\begin{table}[htbp]
\caption{
Systematic uncertainties on \afb, in absolute \%.
For the 2D measurement, the range of changes in \afb\ over the six \mtt\ bins is given.
  \label{tab:sysAfb}
}
\begin{ruledtabular} 
\begin{tabular}{lccc}
 & \head{Reco. level}& \multihead{2} {Production level} \\
\head{Source} & \head{inclusive}& \head{inclusive} & \head{2D} \\
\hline
\tstrut Background model & $+0.7$/$-0.8$ & $1.0$ & $1.1$--$2.8$\\
\tstrut Signal model     & $<0.1$ & $0.5$ & $0.8$--$5.2$\\
\tstrut Unfolding               & N/A &$0.5$ & $0.9$--$1.9$\\
\tstrut PDFs and pileup   &  $0.3$ & $0.4$ & $0.5$--$2.9$\\
\tstrut Detector model   & $+0.1$/$-0.3$ & $0.3$ & $0.4$--$3.3$\\
\tstrut Sample composition  & $<0.1$ & $<0.1$ & $<0.1$\\
\hline                                                                               
\tstrut Total               & $+0.8$/$-0.9$ & $1.3$ & $2.1$--$7.5$\\
\end{tabular}                                                                        
\end{ruledtabular}                                                                         
\end{table}                  
                                                        
The {\bf background model} category includes the following sources, which affect the properties predicted for background events.
The leptonic asymmetry of the \wpj\ background is varied within its uncertainty of $3\%$~\cite{bib:our_afbl}.
The rate of heavy-flavor production within \wpj\ production is varied by $\pm 20\%$~\cite{bib:D0xsect,bib:mcfm}.
The efficiencies for lepton identification, and the probabilities for a jet to be misidentified as a lepton, 
taken as functions of lepton momentum, are varied within their uncertainties to account for the uncertainty on the 
number of background events from multijet production~\cite{bib:matrix_method}. 
This variation affects both the background shape and normalization.
Uncertainties associated with the modeling of the  discriminant, $D_c$, transverse momentum of $W$ boson and \mjjmin, as well as 
 potentially increased background levels at high lepton pseudorapidity
 are also quantified by modifying the background model~\cite{bib:our_afbl}.

The {\bf signal model} category includes the sources of uncertainty that affect the properties predicted for signal events other than the ones accounted for in the  PDFs and pileup category.
The top quark mass is varied according to the 
combined Tevatron measurement of Ref.~\cite{bib:mtop}.
The effect of higher order corrections to \ttbar\ production is estimated by replacing the migration matrix \migmat\ from Eq.~\ref{eq:exp_data} 
simulated by \mcatnlo\ with the one simulated by  \alpgen, which uses tree-level matrix elements.  
The $b$ quark fragmentation function is varied within its uncertainties~\cite{bib:mtop}, which also affects background modeling.

The signal model category also includes the uncertainties associated with gluon radiation.
The total amount of initial state radiation is varied in a range consistent with the results of Ref.~\cite{bib:Zisr}.
We also consider the difference in
the predicted amount of  initial state radiation between forward and backward events, both because
of contributions at order $\alpha_s^3$ and due to higher order effects which are modeled by
the simulated parton showers~\cite{bib:winter}.
We account for this uncertainty by reducing the difference in the distributions of the $p_T$ of the \ttbar\ system for 
forward and backward events  by 25\%, a value derived from Ref.~\cite{bib:winter}.
We also  account for the possibility that the mismodeling of this variable in the \lptj\ final state affects \afb\
by reweighting this distribution to match the \DZ\ data, similarly to the procedure used in Ref.~\cite{bib:our_afbl}.

The uncertainties due to {\bf unfolding} are dominated by the calibration uncertainties. The uncertainties 
associated with the choice of the regularization strength and  statistical fluctuations in the MC samples used 
to find the migration matrix are also included. 

The {\bf PDFs and pileup} category includes uncertainties on the modeling of the \ppbar\ collisions.
The main uncertainties are from the PDFs, which primarily affect  the \dy\ distribution of the \wpj\ background.
These uncertainties are evaluated by varying the contributions of the various eigenvectors 
from the \cteq\ PDF~\cite{bib:CTEQ} and by considering an alternative set of PDFs (\mrst~\cite{bib:MRST}).
The number of additional collisions within the same bunch crossing (pileup) affects the quality of the
reconstruction.  The uncertainties on the modeling of additional collisions are also included in this category.

The {\bf detector model} category includes the following sources of systematic uncertainty.
The efficiencies of the $b$-tagging algorithm for jets of different flavors, which are measured from collider data, 
are varied according to their uncertainties~\cite{bib:btagging}. 
These variations affect the measured \afb\ mostly through the estimated sample composition,
which depends strongly on the classification of data into several channels according to the number of $b$ tags.
The modeling of jet energy reconstruction, including the overall energy scale and the energy resolution, 
as well as jet-reconstruction efficiencies and single-particle responses, are all calibrated to collider data 
and are varied according to their uncertainties~\cite{bib:Jets}.
The uncertainties due to jet reconstruction and energy measurement are significantly reduced
compared to the previous measurement due to the inclusion of the \lptj\ events.

Lastly, the {\bf sample composition} is varied according to its fitted uncertainties.
This variation is performed in addition to the changes in the sample composition implicitly induced by 
other systematic variations.

%
%
\section{Results}
\label{sec:res}

%
\subsection{Inclusive \afb\ and \afb\ dependence on \absdy}
\label{sec:res_dy}
The calibrated \prodhlvl\ \dy\ distribution is shown in Fig.~\ref{fig:dy_dist}.
The corresponding inclusive forward--backward asymmetry in \ttbar\ production is $\left(10.6\pm3.0\right)\%$.

\begin{figure}[htbp]
\begin{center}
\includegraphics[width=0.95\linewidth]{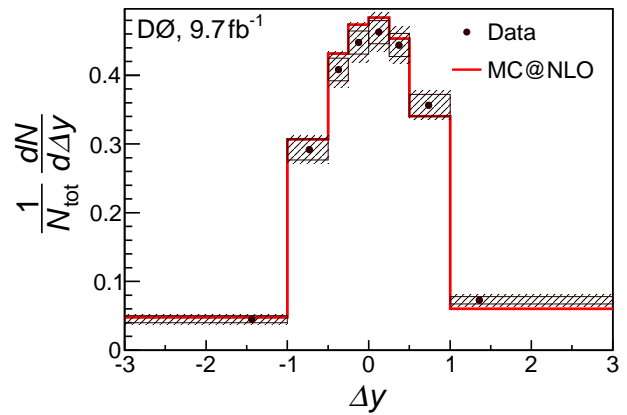}\\
\end{center}
\vspace{-0.6cm}
\caption[Results of the \afb\ by \absdy\ measurement.]{
(Color online). The \prodhlvl\ \dy\ distribution.
The D0 data points are shown with their statistical uncertainty indicated by the black rectangles
and their total uncertainty, based on the diagonal elements of the covariance matrix, indicated by the hashed areas.
The histogram shows the \mcatnlo\ prediction~\cite{bib:mcatnlo}. 
The $x$ coordinate of each data point is the observed average of the \dy\ distribution in the corresponding bin.
}
\label{fig:dy_dist}
\end{figure}

\begin{figure}[htbp]
\begin{center}
\includegraphics[width=0.95\linewidth]{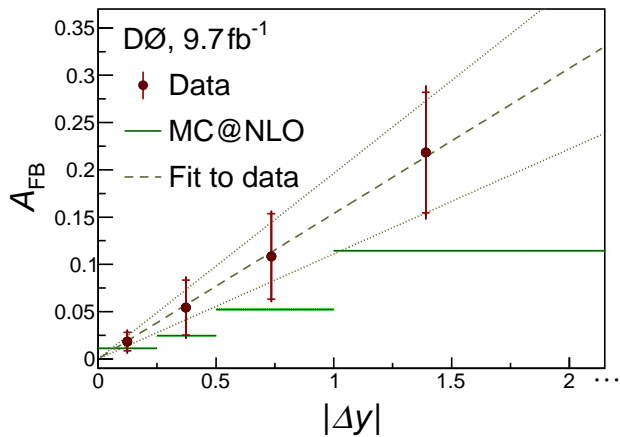}\\
\end{center}
\vspace{-0.6cm}
\caption[Results of the \aady\ measurement.]{
(Color online). The dependence of the forward--backward asymmetry on \absdy.
 The D0 data points are shown with the total error bars indicating the total uncertainty, 
based on the diagonal elements of the covariance matrix, while 
the statistical uncertainties are indicated by the inner error bars.
The dashed line shows the fit to the data with the dotted lines indicating the fit uncertainty.
The horizontal lines show the \mcatnlo\ prediction for the asymmetry in each \mtt\ bin~\cite{bib:mcatnlo}. 
The last bin has no upper boundary.
The $x$ coordinate of each data point is the observed average of \absdy\ in the corresponding bin.
}
\label{fig:res_dy}
\end{figure}

\begin{table}[htbp]
\caption[Calibrated \aady s with their total errors and \mcatnlo\ predictions.]
{Variation of the \prodhlvl\ \afb\ on \absdy.
 The measured values are calibrated and listed with their total uncertainties. 
The theoretical predictions are based on \mcatnlo\ simulation. 
}
\begin{ruledtabular} 
\begin{tabular}{ccc}
                  & \multihead{2}{\afb,\%}\\
\head{\absdy} & \head{Predicted} & \head{Measured}\\
\hline
$<0.25$   & \noone  1.1 & \noone$  1.8 \pm 1.3 $\\
0.25--0.5 & \noone  2.5 & \noone$  5.4 \pm 3.3 $\\
0.5--1    & \noone  5.2 &       $ 10.8 \pm 4.8 $\\
$>1$      &        11.4 &       $ 21.8 \pm 7.1 $\\
\end{tabular}
\end{ruledtabular}
\label{tab:res_dy}
\end{table}

\begin{table}[htbp]
\caption[Correlation factors for \aady\ results.]
{The correlation factors between the measured \afb\ values in different \absdy\ bins.}
\begin{ruledtabular} 
\begin{tabular}{c>{\hspace{6pt}}*{4}{c}}
  & \multihead{4}{\absdy\ range} \\
  & \head{$<0.25$} & \head{0.25--0.5} & \head{0.5--1} & \head{$>1$} \\
  \hline
\head{$<0.25$}   & $+1.00$ & $+0.79$ & $+0.77$ & $+0.06$ \\
\head{0.25--0.5} & $+0.79$ & $+1.00$ & $+0.89$ & $+0.09$ \\
\head{0.5--1}    & $+0.77$ & $+0.89$ & $+1.00$ & $+0.25$ \\
\head{$>1$}      & $+0.06$ & $+0.09$ & $+0.25$ & $+1.00$ \\
\end{tabular}
\end{ruledtabular}
\label{tab:corr_dy}
\end{table}

The dependence of \afb\ on \absdy\ is shown in Fig.~\ref{fig:res_dy} and Table~\ref{tab:res_dy} with the correlation factors between bins listed in Table~\ref{tab:corr_dy}.
These correlations are taken into account in the fit of the measured \aady\ to a line.
Since for any physical \dy\ distribution the asymmetry at $\dy=0$ is 0,
we constrain the line to the origin and fit for its slope.
For data, we find a slope of $0.154\pm0.043$.
This slope is compatible within two standard deviations with the \mcatnlo-simulated slope of $0.080$, 
which  has negligible statistical uncertainty.
The difference between the slope reported by the CDF Collaboration~\cite{bib:CDF2012} 
and the slope reported in this \this\ corresponds to $1.3$ standard deviations\footnote{When comparing to CDF results, 
we neglect the correlations of the systematic uncertainties between the two experiments.}.
 
%
\subsection{\afb\ dependence on \mtt}
\label{sec:res_mtt}
The dependence of \afb\ on \mtt\ is shown in Fig.~\ref{fig:res_2d} and Table~\ref{tab:res} with the correlation factors between bins listed in Table~\ref{tab:corr_2d}.

\begin{table}[htbp]
\caption[Calibrated \afb s with their total errors and \mcatnlo\ predictions.]
{\Prodhlvl\ asymmetries as a function of \mtt. The measured values are calibrated and listed with their total uncertainties. 
The theoretical predictions are based on \mcatnlo\ simulation. 
}
\begin{ruledtabular} 
\begin{tabular}{lcc}
                  & \multihead{2}{\afb,\%}\\
\head{\mtt, \GeV} & \head{Predicted} & \head{Measured}\\
\hline
$<400$       &\noone$2.2$ & \hskip 8.3pt$  7.0\pm  5.1 $\\
$400$--$450$ &\noone$4.6$ & \hskip 8.3pt$  9.3\pm  5.0 $\\
$450$--$500$ &\noone$6.7$ & \hskip 3.4pt$ 12.7\pm  5.7 $\\
$500$--$550$ &\noone$8.4$ & \hskip 3.4pt$ 16.6\pm  8.2 $\\
$550$--$650$ &    $ 10.9$ & \hskip 8.9pt$ 37.6\pm 19.0 $\\
$>650$       &    $ 14.8$ &             $-12.3\pm 29.6 $\\
\hline
Inclusive    &\noone$5.0$ & \hskip 3.1pt$ 10.6\pm  3.0 $\\
\end{tabular}
\end{ruledtabular}
\label{tab:res}
\end{table}
\begin{table}[htbp]
\caption[Correlation factors for 2D results.]
{The correlation factors between the measured \afb\ values in different \mtt\ bins. All masses are in GeV.}
\begin{ruledtabular} 
\begin{tabular}{>{\hspace{6pt}}l*{6}{c}}
  & \multihead{6}{\mtt\ range (\GeV)} \\
  & \head{$<400$} & \head{$400$--$450$} & \head{$450$--$500$} & 
    \head{$500$--$550$} & \head{$550$--$650$} & \head{$>650$} \\
  \hline
  $<400$       & $+1.00$ & $+0.89$ & $+0.39$ & $-0.19$ & $-0.25$ & $+0.12$ \\
  $400$--$450$ & $+0.89$ & $+1.00$ & $+0.67$ & $+0.10$ & $-0.32$ & $+0.12$ \\
  $450$--$500$ & $+0.39$ & $+0.67$ & $+1.00$ & $+0.68$ & $-0.27$ & $+0.05$ \\
  $500$--$550$ & $-0.19$ & $+0.10$ & $+0.68$ & $+1.00$ & $+0.04$ & $-0.12$ \\
  $550$--$650$ & $-0.25$ & $-0.32$ & $-0.27$ & $+0.04$ & $+1.00$ & $-0.41$ \\
  $>650$       & $+0.12$ & $+0.12$ & $+0.05$ & $-0.12$ & $-0.41$ & $+1.00$ \\
\end{tabular}
\end{ruledtabular}
\label{tab:corr_2d}
\end{table}
\begin{table*}[htbp]
\caption[2D results by covariance matrix eigenvectors.]
{The eigenvectors of the covariance matrix \covmat\ and the result of the 2D measurement $\vec{v}$, in the basis of eigenvectors. 
}
\begin{ruledtabular} 
\begin{tabular}{>{\hspace{6pt}}lc@{\hskip-0.5cm}c@{\hskip-0.5cm}c@{\hskip-0.5cm}c@{\hskip-0.5cm}c@{\hskip-0.5cm}cc}
$i$ & \multihead{6}{Eigenvector $\vec{e_i}$} 
   &  \head{$v_i\pm\sigma_i$}\\
  \hline
$1$ & $(-0.592$ & $+0.770$ & $-0.237$ & $-0.007$ & $+0.004$ & $-0.000)$ & $0.000\pm 0.011$\\
 $2$ &$(+0.434$ & $+0.099$ & $-0.775$ & $+0.448$ & $-0.030$ & $+0.002)$ & $0.004\pm0.021$ \\
$3$ &$(+0.673$ & $+0.591$ & $+0.251$ & $-0.339$ & $+0.138$ & $-0.004)$ & $0.130\pm0.071$\\
$4$ & $(+0.034$ & $+0.192$ & $+0.516$ & $+0.826$ & $+0.104$ & $+0.049)$ & $0.256\pm0.093$ \\
 $5$ &$(-0.076$ & $-0.099$ & $-0.113$ & $-0.040$ & $+0.917$ & $+0.360)$ & $0.265\pm0.166$ \\
$6$ &$(-0.029$ & $-0.030$ & $-0.019$ & $+0.031$ & $+0.359$ & $-0.932)$ & $0.247\pm0.311$  \\
\end{tabular}
\end{ruledtabular}
\label{tab:cov_mat}
\end{table*}
\begin{table}[htbp]
\caption[Linear fit to asymmetry vs mtt]
{Parameters of the fit to Eq.~\ref{eq:slope}.
The theoretical predictions are based on the \mcatnlo\ simulation and have negligible statistical uncertainties.}
\centering
\begin{ruledtabular} 
\begin{tabular}{lcc}
\head{Parameter        } & \head{Predicted       } &\head{Measured       } \\
\hline
Slope, $\alpha$\tstrut & $3.8\cdot10^{-4}$ & \noone$(3.9 \pm 4.4)\cdot10^{-4}$\\
Offset, $A_0$          & $5.3\cdot10^{-2}$ & $(11.9 \pm 3.6)\cdot10^{-2}$\\
\end{tabular}
\end{ruledtabular}
\label{tab:slope}
\end{table}

\begin{figure}[htbp]
\begin{center}
\includegraphics[width=0.95\linewidth]{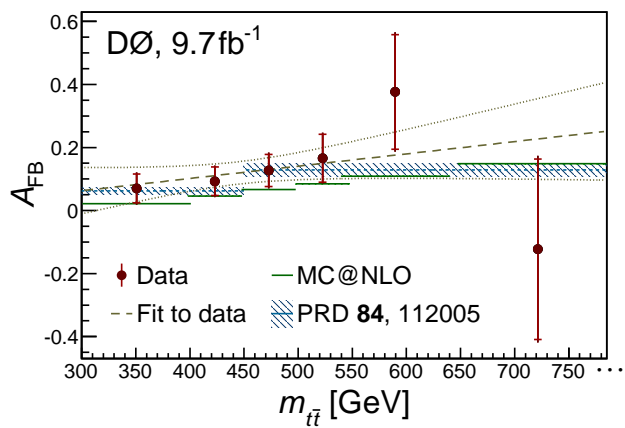}\\
\end{center}
\vspace{-0.6cm}
\caption[Results of the 2D measurement.]{
(Color online). The dependence of the forward--backward asymmetry on the invariant mass of the \ttbar\ system.
The D0 data points are shown with the total error bars indicating the total uncertainty, 
based on the diagonal elements of the covariance matrix, while 
the statistical uncertainties are indicated by the inner error bars.
The dashed line shows a linear fit to the data with the dotted curves indicating the fit uncertainty.
The horizontal lines correspond to the \mcatnlo\  prediction for the asymmetry in each \mtt\ bin~\cite{bib:mcatnlo}. 
Shaded boxes correspond to the prediction of Refs.~\cite{bib:Bern, bib:EWcorr}.
The last bin has no upper boundary.
The $x$ coordinate of each data point is the simulated average of the \mtt\ distribution in the corresponding bin.
}
\label{fig:res_2d}
\end{figure}

The values of the asymmetry measured in six \mtt\  ranges constitute a six-dimensional vector $\vec{v}$ with a $6\times6$ covariance matrix \covmat.
Table~\ref{tab:cov_mat} lists the eigenvectors $\vec{e_i}$ $(i=1,..6)$ of \covmat\  together with the 
 corresponding components of the vector $\vec{v}$ in the basis formed by the eigenvectors: $v_i= \vec{v}\cdot \vec{e_i}$, and their 
 uncertainties $\sigma_i=\sqrt{\boldsymbol{\Sigma'}_{ii}}$, where $\boldsymbol{\Sigma'}$ is the covariance matrix 
transformed to the basis $\vec{e_i}$.
The elements of Table~\ref{tab:cov_mat} fully specify
the measured six-dimensional likelihood in the Gaussian approximation, and can be used for 
quantitative comparison with theoretical predictions and other experimental results~\cite{bib:zech_proc}.  

Using the full covariance matrix we perform a fit of the measured \afb\ 
to the functional form
\begin{equation}
\label{eq:slope}
\afb(\mtt)=\alpha\left(\frac{\mtt}{\textrm{GeV}}-C\right)+A_0.
\end{equation}
We choose $C=445$ so that the correlation factor between the fit parameters $\alpha$ and $A_0$ is  less than $0.01$ in the fit to the data.
The parameters of the fit are listed in Table~\ref{tab:slope} for the data and the \mcatnlo\ simulation. 
We observe a slope $\alpha$ consistent with zero and with the \mcatnlo\ prediction.
The difference between slope reported by the CDF Collaboration~\cite{bib:CDF2012} 
and the slope reported in this \this\ corresponds to $1.8$ standard deviations.

%
%
\section{Discussion}
\label{sec:disc}
The measured inclusive forward--backward asymmetry in \ttbar\ production, $\afb=\left(10.6\pm3.0\right)\%$
is in agreement with the SM predictions reviewed in Section~\ref{sec:Preds}, which range from
an inclusive asymmetry of 5.0\% predicted by the \mcatnlo\ simulation  
to $\left(8.8\pm0.9\right)$\%~\cite{bib:EWcorr} once electroweak effects are taken into account. 
The measured dependences of the asymmetry on \absdy\ and \mtt\ are also in agreement with the SM predictions.
Nevertheless, the observed \afb\ and the dependences of \afb\ on \mtt\ and \absdy\
do not disfavor the larger asymmetries that were previously measured in \ppbar\ collisions~\cite{bib:CDF2012}.

To compare the presented result with the previous D0 publication~\cite{bib:ourPRD},  
Table~\ref{tab:comp} presents \afb\ at the reconstruction level measured in different samples.  
The method discussed in this \this\ applied to \lpgefj\ events from the first 5.4\ifb\ of integrated luminosity yields a result 
consistent with that in Ref.~\cite{bib:ourPRD}, 
but with a reduced uncertainty mainly due to the separation of data into channels based on the number of $b$ tags 
and the increased efficiency of the new $b$-tagging algorithm. 
Once the analysis is extended to include the \lptj\ events collected at that time,
the uncertainty is reduced by a factor of 1.26. 
The result obtained in the second 4.3\ifb\ of the Tevatron dataset is within one standard deviation from  that 
obtained in the first 5.4\ifb. 
The statistical uncertainty obtained in the combined 9.7\ifb\ of integrated luminosity is reduced by a factor of 1.29 with 
respect to the result obtained using the same method in the first 5.4\ifb, 
while the reduction expected from scaling with the integrated luminosity is 1.34. 
This loss of sensitivity is mainly due to higher instantaneous luminosity during the collection of the later data, 
which required a tighter trigger selection. 

\begin{table}[htbp]
\caption[Calibrated \afb s with their total errors and \mcatnlo\ predictions.]
{\Recohlvl\ asymmetries measured in different samples with different methods, with
  their statistical uncertainties. }
\begin{ruledtabular} 
\begin{tabular}{lcc}
& & \head{Reco-level}   \\
\head{Sample} &  \head{Method}  &\head{ \afb, \%}   \\
\hline
 \lpgefj, first 5.4\ifb   & From Ref.~\cite{bib:ourPRD}   &  \noone$9.2 \pm 3.7$\tstrut  \\
 \lpgefj, first 5.4\ifb   & This analysis   &  \noone$9.9 \pm 3.4$   \\
 \lpgetj, first 5.4\ifb   &   This analysis &  $10.1 \pm 2.7$   \\
 \lpgetj, additional 4.3\ifb   & This analysis   &  \noone$6.0 \pm 3.1$   \\
 \lpgetj, full 9.7\ifb   & This analysis   &  \noone$7.9 \pm 2.1$   \\
\end{tabular}
\end{ruledtabular}
\label{tab:comp}
\end{table}

The improved reconstruction of \dy\ and the reduced acceptance bias due to the inclusion of the \lptj\ events result 
in further reduction of the statistical uncertainty on the unfolded result compared to Ref.~\cite{bib:ourPRD}.
The separation of the data into channels allows  us to add the \lptj\ 
channels without losing the statistical power of the purer \lpgefj\ channels. 

%
%
\section{Summary}
\label{sec:concl}

In summary, we report the measurement of the forward--backward asymmetry in \ttbar\ production using the dataset
recorded by the \DZ\ detector in Run II of the Fermilab Tevatron Collider. 
The results presented here supersede the ones that were based on about half of the data~\cite{bib:ourPRD}. 
The analysis is extended to include events with three jets, allowing for the loss of one jet from the \ttbar\ decay
and reducing the acceptance corrections. 
The unfolding procedure now accounts for the differences in sample compositions between channels, thus maximizing the
statistical strength of the individual channels. 
New reconstruction techniques are used in the \lpgefj\ channel, 
improving the experimental resolution in all variables of interest. 

The asymmetry measured at the reconstruction level is $\afb=\left( 7.9 \pm 2.3\right)\%$. 
After correcting for detector resolution and acceptance, we obtain a \prodhlvl\  asymmetry 
$\afb=\left(10.6 \pm  3.0 \right) \%$.
The observed asymmetry and the dependences of \afb\ on \mtt\ and \absdy\
are consistent with the standard model predictions.
\vspace{0.9cm}
\begin{acknowledgments}
We thank M.~Mangano, P.~Skands and G.~Perez for enlightening discussions.
\input{acknowledgement.tex}
\end{acknowledgments}

\end{document}

%% file: author_list_revised.tex
%
\affiliation{LAFEX, Centro Brasileiro de Pesquisas F\'{i}sicas, Rio de Janeiro, Brazil}
\affiliation{Universidade do Estado do Rio de Janeiro, Rio de Janeiro, Brazil}
\affiliation{Universidade Federal do ABC, Santo Andr\'e, Brazil}
\affiliation{University of Science and Technology of China, Hefei, People's Republic of China}
\affiliation{Universidad de los Andes, Bogot\'a, Colombia}
\affiliation{Charles University, Faculty of Mathematics and Physics, Center for Particle Physics, Prague, Czech Republic}
\affiliation{Czech Technical University in Prague, Prague, Czech Republic}
\affiliation{Institute of Physics, Academy of Sciences of the Czech Republic, Prague, Czech Republic}
\affiliation{Universidad San Francisco de Quito, Quito, Ecuador}
\affiliation{LPC, Universit\'e Blaise Pascal, CNRS/IN2P3, Clermont, France}
\affiliation{LPSC, Universit\'e Joseph Fourier Grenoble 1, CNRS/IN2P3, Institut National Polytechnique de Grenoble, Grenoble, France}
\affiliation{CPPM, Aix-Marseille Universit\'e, CNRS/IN2P3, Marseille, France}
\affiliation{LAL, Universit\'e Paris-Sud, CNRS/IN2P3, Orsay, France}
\affiliation{LPNHE, Universit\'es Paris VI and VII, CNRS/IN2P3, Paris, France}
\affiliation{CEA, Irfu, SPP, Saclay, France}
\affiliation{IPHC, Universit\'e de Strasbourg, CNRS/IN2P3, Strasbourg, France}
\affiliation{IPNL, Universit\'e Lyon 1, CNRS/IN2P3, Villeurbanne, France and Universit\'e de Lyon, Lyon, France}
\affiliation{III. Physikalisches Institut A, RWTH Aachen University, Aachen, Germany}
\affiliation{Physikalisches Institut, Universit\"at Freiburg, Freiburg, Germany}
\affiliation{II. Physikalisches Institut, Georg-August-Universit\"at G\"ottingen, G\"ottingen, Germany}
\affiliation{Institut f\"ur Physik, Universit\"at Mainz, Mainz, Germany}
\affiliation{Ludwig-Maximilians-Universit\"at M\"unchen, M\"unchen, Germany}
\affiliation{Panjab University, Chandigarh, India}
\affiliation{Delhi University, Delhi, India}
\affiliation{Tata Institute of Fundamental Research, Mumbai, India}
\affiliation{University College Dublin, Dublin, Ireland}
\affiliation{Korea Detector Laboratory, Korea University, Seoul, Korea}
\affiliation{CINVESTAV, Mexico City, Mexico}
\affiliation{Nikhef, Science Park, Amsterdam, the Netherlands}
\affiliation{Radboud University Nijmegen, Nijmegen, the Netherlands}
\affiliation{Joint Institute for Nuclear Research, Dubna, Russia}
\affiliation{Institute for Theoretical and Experimental Physics, Moscow, Russia}
\affiliation{Moscow State University, Moscow, Russia}
\affiliation{Institute for High Energy Physics, Protvino, Russia}
\affiliation{Petersburg Nuclear Physics Institute, St. Petersburg, Russia}
\affiliation{Instituci\'{o} Catalana de Recerca i Estudis Avan\c{c}ats (ICREA) and Institut de F\'{i}sica d'Altes Energies (IFAE), Barcelona, Spain}
\affiliation{Uppsala University, Uppsala, Sweden}
\affiliation{Taras Shevchenko National University of Kyiv, Kiev, Ukraine}
\affiliation{Lancaster University, Lancaster LA1 4YB, United Kingdom}
\affiliation{Imperial College London, London SW7 2AZ, United Kingdom}
\affiliation{The University of Manchester, Manchester M13 9PL, United Kingdom}
\affiliation{University of Arizona, Tucson, Arizona 85721, USA}
\affiliation{University of California Riverside, Riverside, California 92521, USA}
\affiliation{Florida State University, Tallahassee, Florida 32306, USA}
\affiliation{Fermi National Accelerator Laboratory, Batavia, Illinois 60510, USA}
\affiliation{University of Illinois at Chicago, Chicago, Illinois 60607, USA}
\affiliation{Northern Illinois University, DeKalb, Illinois 60115, USA}
\affiliation{Northwestern University, Evanston, Illinois 60208, USA}
\affiliation{Indiana University, Bloomington, Indiana 47405, USA}
\affiliation{Purdue University Calumet, Hammond, Indiana 46323, USA}
\affiliation{University of Notre Dame, Notre Dame, Indiana 46556, USA}
\affiliation{Iowa State University, Ames, Iowa 50011, USA}
\affiliation{University of Kansas, Lawrence, Kansas 66045, USA}
\affiliation{Louisiana Tech University, Ruston, Louisiana 71272, USA}
\affiliation{Northeastern University, Boston, Massachusetts 02115, USA}
\affiliation{University of Michigan, Ann Arbor, Michigan 48109, USA}
\affiliation{Michigan State University, East Lansing, Michigan 48824, USA}
\affiliation{University of Mississippi, University, Mississippi 38677, USA}
\affiliation{University of Nebraska, Lincoln, Nebraska 68588, USA}
\affiliation{Rutgers University, Piscataway, New Jersey 08855, USA}
\affiliation{Princeton University, Princeton, New Jersey 08544, USA}
\affiliation{State University of New York, Buffalo, New York 14260, USA}
\affiliation{University of Rochester, Rochester, New York 14627, USA}
\affiliation{State University of New York, Stony Brook, New York 11794, USA}
\affiliation{Brookhaven National Laboratory, Upton, New York 11973, USA}
\affiliation{Langston University, Langston, Oklahoma 73050, USA}
\affiliation{University of Oklahoma, Norman, Oklahoma 73019, USA}
\affiliation{Oklahoma State University, Stillwater, Oklahoma 74078, USA}
\affiliation{Brown University, Providence, Rhode Island 02912, USA}
\affiliation{University of Texas, Arlington, Texas 76019, USA}
\affiliation{Southern Methodist University, Dallas, Texas 75275, USA}
\affiliation{Rice University, Houston, Texas 77005, USA}
\affiliation{University of Virginia, Charlottesville, Virginia 22904, USA}
\affiliation{University of Washington, Seattle, Washington 98195, USA}
\author{V.M.~Abazov} \affiliation{Joint Institute for Nuclear Research, Dubna, Russia}
\author{B.~Abbott} \affiliation{University of Oklahoma, Norman, Oklahoma 73019, USA}
\author{B.S.~Acharya} \affiliation{Tata Institute of Fundamental Research, Mumbai, India}
\author{M.~Adams} \affiliation{University of Illinois at Chicago, Chicago, Illinois 60607, USA}
\author{T.~Adams} \affiliation{Florida State University, Tallahassee, Florida 32306, USA}
\author{J.P.~Agnew} \affiliation{The University of Manchester, Manchester M13 9PL, United Kingdom}
\author{G.D.~Alexeev} \affiliation{Joint Institute for Nuclear Research, Dubna, Russia}
\author{G.~Alkhazov} \affiliation{Petersburg Nuclear Physics Institute, St. Petersburg, Russia}
\author{A.~Alton$^{a}$} \affiliation{University of Michigan, Ann Arbor, Michigan 48109, USA}
\author{A.~Askew} \affiliation{Florida State University, Tallahassee, Florida 32306, USA}
% ...
%
%
\collaboration{The D0 Collaboration\footnote{with visitors from
$^{a}$Augustana College, Sioux Falls, SD, USA,
$^{b}$The University of Liverpool, Liverpool, UK,
$^{c}$DESY, Hamburg, Germany,
$^{d}$Universidad Michoacana de San Nicolas de Hidalgo, Morelia, Mexico
$^{e}$SLAC, Menlo Park, CA, USA,
$^{f}$University College London, London, UK,
$^{g}$Centro de Investigacion en Computacion - IPN, Mexico City, Mexico,
$^{h}$Universidade Estadual Paulista, S\~ao Paulo, Brazil,
$^{i}$Karlsruher Institut f\"ur Technologie (KIT) - Steinbuch Centre for Computing (SCC),
D-76128 Karlsruhe, Germany,
$^{j}$Office of Science, U.S. Department of Energy, Washington, D.C. 20585, USA,
$^{k}$American Association for the Advancement of Science, Washington, D.C. 20005, USA,
$^{l}$Kiev Institute for Nuclear Research, Kiev, Ukraine,
$^{m}$University of Maryland, College Park, Maryland 20742, USA,
and
$^{n}$Laboratoire de Physique Theorique, Orsay, FR.
}} \noaffiliation
\vskip 0.25cm

%% file: acknowledgement.tex
%
We thank the staffs at Fermilab and collaborating institutions,
and acknowledge support from the
DOE and NSF (USA);
CEA and CNRS/IN2P3 (France);
MON, NRC KI and RFBR (Russia);
CNPq, FAPERJ, FAPESP and FUNDUNESP (Brazil);
DAE and DST (India);
Colciencias (Colombia);
CONACyT (Mexico);
NRF (Korea);
FOM (The Netherlands);
STFC and the Royal Society (United Kingdom);
MSMT and GACR (Czech Republic);
BMBF and DFG (Germany);
SFI (Ireland);
The Swedish Research Council (Sweden);
and
CAS and CNSF (China).